\documentclass[12pt]{iopart}
\usepackage{graphicx,bbm,amsthm,amsfonts,amssymb,amsopn,amsmath}
\usepackage{dsfont}
\usepackage{mathtools}
\usepackage{cite}
\usepackage{hyperref}
\usepackage{pgf,tikz}
\usepackage{iopams}
\usetikzlibrary{arrows,calc,matrix,backgrounds,fit,positioning}
\usepackage{calc}
\usepackage{xstring}
\usepackage{ifthen}
\usepackage{pdftexcmds}
\usepackage{adjustbox}

\def\be{\begin{equation}}
\def\ee{\end{equation}}
\def\bra#1{\mathinner{\langle{#1}|}}
\def\ket#1{\mathinner{|{#1}\rangle}}

\def\p{{\!\rm p}}

\def\one{\mathbbm{1}}

\theoremstyle{theorem}

\newtheorem{thm}{Theorem}
\theoremstyle{remark}
\newtheorem*{rem}{Remark}
\newcommand{\bea}{\begin{eqnarray}}
\newcommand{\eea}{\end{eqnarray}}

\newcommand{\ave}[1]{{\langle #1\rangle}}

\newcommand{\ii}{ {\rm i} }

\newcommand{\ZZ}{\mathbb{Z}}

\newcommand{\x}{{\rm x}}

\newcommand{\oo}[1]{{[ #1 ]}}

\newcommand{\lsp}{\text{lsp}}

\newcommand{\erf}{{\mathrm{erf}}}
\renewcommand{\vec}[1]{{\mathbf{#1}}}

\def\tit#1{}

\eqnobysec

\usepackage{bbm}

\definecolor{full}{rgb}{0,0,0}
\definecolor{old}{rgb}{1,1,1}
\definecolor{half}{rgb}{0.9,0.9,0.9}
\definecolor{border}{rgb}{0.3,0.3,0.3}
\definecolor{halfborder}{rgb}{0.5,0.5,0.5}
\def\a{0.35}
\def\b{8}
\def\off{1.75}
\def\num{3}
\newcommand\emptyrectangle[2]{
  \draw[thick,border] ({\a*(#1)},{\a*(#2-1)})  -- ({\a*(#1+1)},{\a*(#2)})  -- ({\a*(#1)},{\a*(#2+1)})  
  -- ({\a*(#1-1)},{\a*(#2)})  -- cycle;
}
\newcommand\fullrectangle[2]{
  \draw[thick,border,fill=full] ({\a*(#1)},{\a*(#2-1)})  -- ({\a*(#1+1)},{\a*(#2)})  -- ({\a*(#1)},{\a*(#2+1)})  
  -- ({\a*(#1-1)},{\a*(#2)})  -- cycle;
}
\newcommand\halfrectangle[2]{
  \draw[thick,dashed,halfborder,fill=half] ({\a*(#1)},{\a*(#2-1)})  -- ({\a*(#1+1)},{\a*(#2)})  -- ({\a*(#1)},{\a*(#2+1)})  
  -- ({\a*(#1-1)},{\a*(#2)})  -- cycle;
}
\newcommand\unknownrectangle[2]{
  \draw[thin,dashed,border] ({\a*(#1)},{\a*(#2-1)})  -- ({\a*(#1+1)},{\a*(#2)})  -- ({\a*(#1)},{\a*(#2+1)})  
  -- ({\a*(#1-1)},{\a*(#2)})  -- cycle;
}
\newcommand\nrectangle[3]{
  \draw[black] ({\a*#1-(0.5*\x)},{\a*#2-(0.5*\x)}) rectangle 
  ({\a*#1+(0.5*\x)},{\a*#2+(0.5*\x)});
  \node[font=\small] at ({\a*#1},{\a*#2+0.5*\x+\off}) {#3};
}
\newcommand\brectangle[1]{
  \draw[very thick,gray,rounded corners] ({-\a*#1-0.5*\x-\off},{0.5*\x+2*\off}) rectangle
  ({-\a*(#1-0.5)+0.5*\x+\off},{-0.5*\a-0.5*\x-\off});
}

\begin{document}

\title{Time-dependent matrix product ansatz for interacting reversible dynamics}

\author{Katja Klobas, Marko Medenjak, Toma\v z Prosen, Matthieu Vanicat}
\address{Department of Physics, Faculty of Mathematics and Physics, University of Ljubljana, Ljubljana, Slovenia}

\begin{abstract}
We present an explicit time-dependent matrix product ansatz (tMPA) which
describes the time-evolution of any local observable in an interacting and
deterministic lattice gas, specifically 
for the rule 54 reversible cellular automaton  of [Bobenko {\em et al}. Commun.
Math. Phys. {\bf 158}, 127 (1993)]. 
Our construction is based on an explicit solution of real-space real-time inverse scattering problem.
We consider two applications of this tMPA. 
Firstly, we provide the first exact and explicit computation of the dynamic structure factor in an interacting deterministic model,
and secondly, we solve the extremal case of the inhomogeneous quench problem,
where a semi-infinite lattice in the maximum entropy state is joined with an
empty semi-infinite lattice.
Both of these exact results rigorously demonstrate a coexistence of ballistic
and diffusive transport behaviour in the model, as expected for normal fluids.
\end{abstract}

\section{Introduction}

Understanding rigorously how the macroscopic hydrodynamic behaviour of
interacting particle systems emerges from the microscopic description is one of
the major quests of nonequilibrium statistical mechanics~\cite{Spohn}. This
problem is particularly hard in systems, where the microscopic equations of
motion are reversible and no external sources of noise or dissipation are built
into the model.  The standard route of deriving the macroscopic description for
the interacting systems with hyperbolic (i.e.\ `chaotic') microscopic dynamics,
say for elastically colliding hard-spheres, goes via the justification of the
Boltzmann equation, which is possible in the Boltzmann-Grad
limit~\cite{Golse2013}. Despite this limitation, the hydrodynamic approach has been
recently heuristically demonstrated even in quantum integrable
systems~\cite{Doyon2016,Bertini2016,Doyon2018,Jacopo2018} where any possible mechanism of chaos is absent.

In this paper we discuss an interacting deterministic many-body classical
system in a single spatial dimension, for which we can explicitly compute the
complete dynamics of {\em all} local observables.  This is a reversible
cellular automaton given by the rule $54$ (RCA $ 54 $) of Bobenko \emph{et
al}~\cite{bobenko1993two}, also related to a model coded as ERCA 250R
introduced by Takesue~\cite{Takesue}.  The model is a two state
locally interacting deterministic lattice system that describes the dynamics of
classical solitons with nontrivial pairwise scattering.  Recently
non-equilibrium steady state of the system coupled to stochastic reservoirs has
been found~\cite{prosenMejiaMonasterioCA54,inoueTakesueTwoExtensionsCA54}, as
well as some of its decay modes~\cite{prosenBucaDecayModesCA54}. These results
suggest that the model can be indeed considered as an integrable Hamiltonian
system although no Lax zero curvature formulation of the equations of motion
has been found to date.  The model has as well been studied in the quantum
context, since it can also be interpreted as a quantum cellular automaton
describing spreading of time dependent local
operators~\cite{gopalakrishnanOperatorGrowth}. 

The key new concept in our work is an exact and explicit time dependent matrix
product anastz (tMPA) representation of time-evolution of local observables.
The dimension of the auxiliary space which supports the matrix representation
is formally infinite, but in time $t$ it only explores a polynomial, in fact
${\cal O}(t^2)$, dimensional subspace, which implies efficient computation of
dynamics. As an application of our technique we provide explicit and large
time/space asymptotic results of the dynamic structure factor (i.e. spatio-temporal
density-density correlation function in the maximum entropy
equilibrium state) as well as an explicit solution of the inhomogeneous quench
problem of joining a maximum entropy semi-infinite lattice with an empty
semi-infinite lattice. Both explicit solutions demonstrate a coexistence of
ballistic (convective) and diffusive (conductive) transport, which is typically
to be expected in normal gasses or liquids.

The paper is structured as follows: In section~\ref{sec:model} we describe the
dynamics of the RCA $54$. In section~\ref{sec:tMPA} we present the main
result; the time propagation of the local observables is expressed as a tMPA,
which is explicitly derived by solving a real-time real-space inverse
scattering problem.  Despite the infinite dimensionality of the tMPA, we are
able to obtain the solution of two physically interesting problems by
exploiting the structure of the matrices (as explained in~\ref{sec:actionOfT}).
In section~\ref{sec:inhQ} we solve the inhomogeneous quench problem, where the
left-hand side of the system is prepared in the maximally mixed state, and the
right-hand side is completely vacant. The second problem we address is the
analytical calculation of the dynamic structure factor, presented in
section~\ref{sec:Cxt}.

\section{The model}\label{sec:model}

\subsection{Definition of the dynamics}
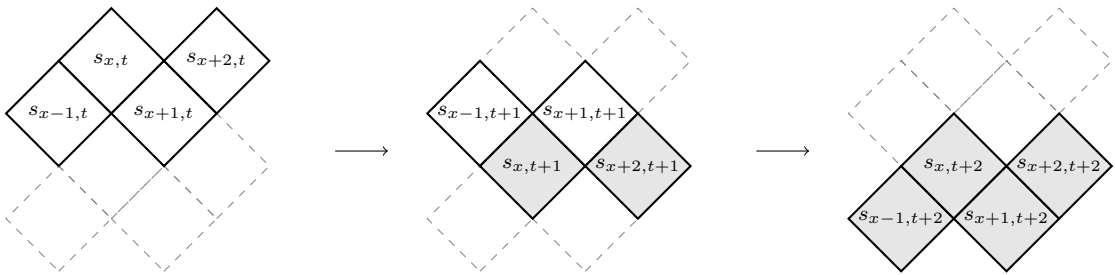
\begin{figure}
	\centering
	\definecolor{full}{rgb}{0,0,0}
\definecolor{old}{rgb}{1,1,1}
\definecolor{updated}{rgb}{0.9,0.9,0.9}
\definecolor{border}{rgb}{0.5,0.5,0.5}
\begin{tikzpicture}
  \def\a{0.7}
  \def\b{8}
  \def\off{1.75}
  \newcommand\coloredrectangle[3]{
    \draw[thick,black,fill=#3] ({\a*(#1)},{\a*(#2-1)})  -- ({\a*(#1+1)},{\a*(#2)})  -- ({\a*(#1)},{\a*(#2+1)})  
    -- ({\a*(#1-1)},{\a*(#2)})  -- cycle;
  }
  \newcommand\nbrectangle[2]{
    \draw[very thin,dashed,border] ({\a*(#1)},{\a*(#2-1)})  -- ({\a*(#1+1)},{\a*(#2)})  -- ({\a*(#1)},{\a*(#2+1)})  
    -- ({\a*(#1-1)},{\a*(#2)})  -- cycle;
  }
  \newcommand\bsawup[2]{
    \nbrectangle{#1+0}{#2+0};
    \nbrectangle{#1+1}{#2+1};
    \nbrectangle{#1+2}{#2+0};
    \nbrectangle{#1+3}{#2+1};
  }
  \newcommand\bsawdown[2]{
    \nbrectangle{#1+0}{#2+0};
    \nbrectangle{#1+1}{#2-1};
    \nbrectangle{#1+2}{#2+0};
    \nbrectangle{#1+3}{#2-1};
  }
  \newcommand\sawup[6]{
    \coloredrectangle{#1+0}{#2+0}{#3};
    \coloredrectangle{#1+1}{#2+1}{#4};
    \coloredrectangle{#1+2}{#2+0}{#5};
    \coloredrectangle{#1+3}{#2+1}{#6};
  }
  \newcommand\sawdown[6]{
    \coloredrectangle{#1+0}{#2+0}{#3};
    \coloredrectangle{#1+1}{#2-1}{#4};
    \coloredrectangle{#1+2}{#2+0}{#5};
    \coloredrectangle{#1+3}{#2-1}{#6};
  }
  \newcommand\uplabels[6]{
    \node at ({\a*(#1)},{\a*(#2)}) {#3};
    \node at ({\a*(#1+1)},{\a*(#2+1)}) {#4};
    \node at ({\a*(#1+2)},{\a*(#2)}) {#5};
    \node at ({\a*(#1+3)},{\a*(#2+1)}) {#6};
  }
  \newcommand\downlabels[6]{
    \node at ({\a*(#1)},{\a*(#2)}) {#3};
    \node at ({\a*(#1+1)},{\a*(#2-1)}) {#4};
    \node at ({\a*(#1+2)},{\a*(#2)}) {#5};
    \node at ({\a*(#1+3)},{\a*(#2-1)}) {#6};
  }

  \bsawup{0}{-2};
  \sawup{0}{0}{old}{old}{old}{old};
  \uplabels{0}{0}{${\scriptstyle s_{x-1,t}}$}{${\scriptstyle s_{x,t}}$}{${\scriptstyle s_{x+1,t}}$}{${\scriptstyle s_{x+2,t}}$};

  \bsawup{\b}{0};
  \bsawup{\b}{-2};
  \sawdown{\b}{0}{old}{updated}{old}{updated};
  \downlabels{\b}{0}{${\scriptstyle s_{x-1,t+1}}$}{${\scriptstyle s_{x,t+1}}$}{${\scriptstyle s_{x+1,t+1}}$}{${\scriptstyle s_{x+2,t+1}}$};

  \bsawup{2*\b}{0};
  \sawup{2*\b}{-2}{updated}{updated}{updated}{updated};
  \uplabels{2*\b}{-2}{${\scriptstyle s_{x-1,t+2}}$}{${\scriptstyle s_{x,t+2}}$}{${\scriptstyle s_{x+1,t+2}}$}{${\scriptstyle s_{x+2,t+2}}$};
  \draw[->] ({3.5*\a+\off*\a},-0.5) -- ({\b*\a-\off*\a},{-0.5});
  \draw[->] ({\b*\a+3.5*\a+\off*\a},-0.5) -- ({2*\b*\a-\off*\a},{-0.5});
\end{tikzpicture}
	\caption{\label{fig:LatticeGeometry} Schematic representation of the time evolution of a section of the lattice.
		In the first time-step, only the sites $x$ and $x+2$ are updated, while the states on the sites
		$x-1$ and $x+1$ remain unchanged. In the second time step, the sites $x\pm 1$ are updated,
		and $x$, $x+2$ do not change.
	}
\end{figure}

\begin{figure}
  \centering
  \definecolor{full}{rgb}{0,0,0}
\definecolor{empty}{rgb}{1,1,1}
\definecolor{border}{rgb}{0.3,0.3,0.3}
\begin{tikzpicture}
  \def\a{0.43}
  \def\b{4.25}
  \newcommand\labels[2]{
    \node[border] at ({\a*(-1+#1)},{\a*(#2)}) {$s_1$};
    \node[border] at ({\a*(#1)},{\a*(1+#2)}) {$s_2$};
    \node[border] at ({\a*(1+#1)},{\a*(#2)}) {$s_3$};
    \node[border] at ({\a*(#1)},{\a*(-1+#2)}) {$s_2^{\prime}$};
  }
  \newcommand\coloredrectangle[3]{
    \draw[border,fill=#3] ({\a*(#1)},{\a*(#2-1)})  -- ({\a*(#1+1)},{\a*(#2)})  -- ({\a*(#1)},{\a*(#2+1)})  
    -- ({\a*(#1-1)},{\a*(#2)})  -- cycle;
  }
  \newcommand\dcoloredrectangle[3]{
    \draw[thick,red,fill=#3] ({\a*(#1)},{\a*(#2-1)})  -- ({\a*(#1+1)},{\a*(#2)})  -- ({\a*(#1)},{\a*(#2+1)})  
    -- ({\a*(#1-1)},{\a*(#2)})  -- cycle;
  }
  \newcommand\onerule[6]{
    \coloredrectangle{-1+#1}{#2}{#3};
    \coloredrectangle{#1}{1+#2}{#4};
    \coloredrectangle{1+#1}{#2}{#5};
    \dcoloredrectangle{#1}{-1+#2}{#6};
  }
  \onerule{0}{0}{empty}{empty}{empty}{empty};
  \onerule{\b}{0}{empty}{empty}{full}{full};
  \onerule{2*\b}{0}{empty}{full}{empty}{full};
  \onerule{3*\b}{0}{empty}{full}{full}{empty};
  \onerule{4*\b}{0}{full}{empty}{empty}{full};
  \onerule{5*\b}{0}{full}{empty}{full}{full};
  \onerule{6*\b}{0}{full}{full}{empty}{empty};
  \onerule{7*\b}{0}{full}{full}{full}{empty};
  \labels{0}{0};
\end{tikzpicture}
  \caption{\label{fig:54Rules}In the figure, the RCA $54$ is presented diagrammatically.
    The updated value $s_2^{\prime}$, i.e. the square with a red-border,
    depends on the values $s_1$, $s_2$, $s_3$ (the top three squares)
    and is given by the local map $s_2^{\prime}=\chi(s_1,s_2,s_3)$
    as defined in~\eqref{eq:54Rules}.
  }
\end{figure}
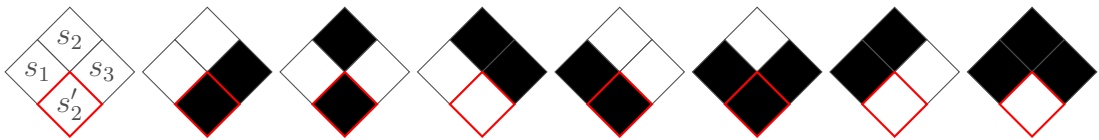

The model is defined on the infinite chain $\mathbb{Z}$,
and each site of the chain can be either occupied or empty.
The configuration
of the system at time $t$ is given by a string of binary digits
$\underline{s}^t=(\ldots,s_{-1}^t,s_{0}^t,s_{1}^t,s_{2}^t,\ldots)\in\{0,1\}^{\mathbb{Z}}$,
where $s_{x}^t=0$, if the site $x$ is empty at time $t$ and $s_{x}^t=1$, if it is
occupied. The states are put on a saw shaped lattice, and the dynamics is provided by the staggered reversible deterministic discrete space-time mapping
\begin{eqnarray}
\label{tp}
  \underline{s}^{t+1} = \begin{cases}
    {\cal M}^{\text{e}} \left(\underline{s}^{t}\right); &t\equiv 0\pmod{2},\\
    {\cal M}^{\text{o}} \left(\underline{s}^{t}\right); &t\equiv 1\pmod{2},
  \end{cases}
\end{eqnarray}
where ${\cal M}^{\text{e}}:\underline{s}\mapsto \underline{s}'$ and ${\cal M}^{\text{o}}:\underline{s}\mapsto \underline{s}''$ 
are maps from $\{0,1\}^{\mathbb{Z}}$ to $\{0,1\}^{\mathbb{Z}}$,  defined by the local three-site updates
\begin{eqnarray}\fl
  s_x' = \begin{cases}
  \chi(s_{x-1},s_{x},s_{x+1}); &\mkern-9mu x\equiv 0 \mkern-16mu \pmod{2}, \\
  s_{x} &\mkern-9mu x\equiv 1 \mkern-16mu \pmod{2},
  \end{cases} \quad
  s_x'' = \begin{cases}
    s_{x}; & \mkern-9mu x\equiv 0 \mkern-16mu \pmod{2}, \\
    \chi(s_{x-1},s_{x},s_{x+1}); &\mkern-9mu x\equiv 1 \mkern-16mu \pmod{2}.
  \end{cases}        
\end{eqnarray}
The schematic representation of the time evolution is presented in Figure \ref{fig:LatticeGeometry}.
The RCA 54 is described by the binary function $ \chi $,
\begin{eqnarray}\label{eq:54Rules}
\chi(s_1,s_2,s_3)=s_1+s_2+s_3+s_1\,s_3\pmod{2}.
\end{eqnarray}
The rules \eqref{eq:54Rules} are diagrammatically expressed in Figure  \ref{fig:54Rules}. 
The complete time evolution is obtained by alternately applying the maps ${\cal M}^{\text{e}}$ and ${\cal M}^{\text{o}}$.
Note that ${\cal M}^{\text{e}}$ and ${\cal M}^{\text{o}}$ encode exactly the same rules shifted by a single lattice site.
The dynamics induced by the mapping \eqref{tp} is time-reversible since the relation
\begin{eqnarray}
\chi(s_1,\chi(s_1,s_2,s_3),s_3)=s_2,
\end{eqnarray}
is satisfied for all $s_1,s_2,s_3\in\{0,1\}$.

\begin{figure}
  \centering
  \includegraphics[width=\linewidth]{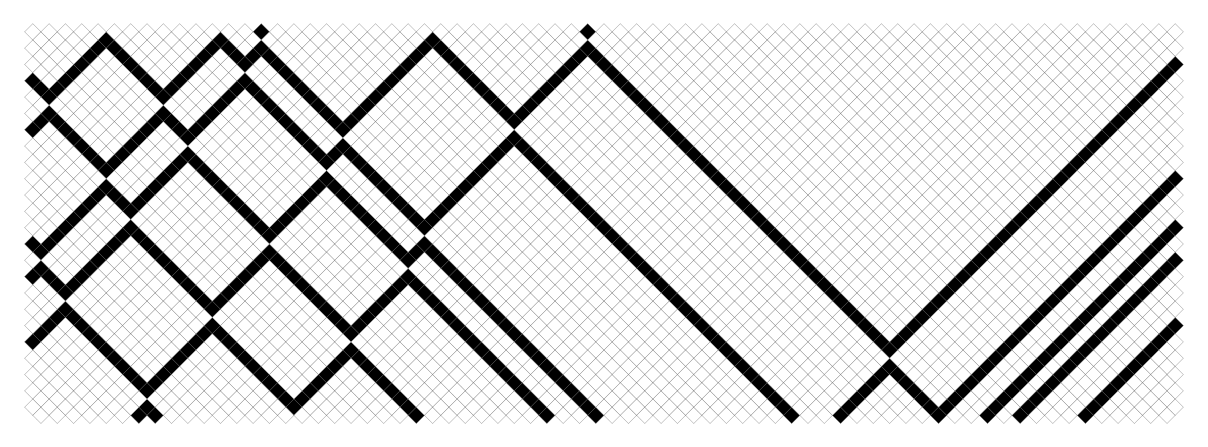}
  \caption{\label{fig:OneRandomConfig} Time evolution of a random initial configuration. Particles move
    with velocity $1$ and scatter pairwise by obtaining a time shift.
  }
\end{figure}

Alternatively, the time propagation \eqref{tp} can be represented by the following prescription
\begin{eqnarray}
  s_{x}^{t+1}=\begin{cases}
    \chi(s_{x-1}^t,s_{x}^t,s_{x+1}^t);& x+t\equiv 0 \pmod{2},\\
    s_{x}^t;& x+t \equiv 1 \pmod{2}.
  \end{cases}
\end{eqnarray}
The physical interpretation of the dynamics induced by the RCA 54 is rather simple.
Occupied sites can be interpreted as the solitons moving with a constant
velocity $\pm 1$, or two scattering solitons, depending on the states of the
neighboring two sites.
After the scattering both solitons
are time-delayed for a single time-step, see Figure \ref{fig:OneRandomConfig}.
The solitons with velocity $\pm 1 $ will be called the left- or right-movers,
respectively.

In the following paragraphs we introduce the necessary prerequisites in order to explicitly study statistical mechanics of the model.

\subsection{Algebra of local observables}
In order to express and  efficiently compute the expectation values of physical
quantities, we introduce a commutative quasi-local $C^*$ algebra $\mathcal{A}$
of observables,
\footnote{${\cal A}$ can be considered as a subalgebra (of diagonal operators,
i.e.\ those jointly commuting with $z$-components of all local spin operators)
of the quasi-local spin $1/2$ UHF algebra \cite{Bratteli}.} or functions over
$\{0,1\}^\mathbb{Z}$. 
 Any local subalgebra ${\cal A}_x \subset {\cal A}$ pertaining to the site $x\in\mathbb{Z}$ is spanned by the local
basis $\oo{\alpha}_x$, $\alpha\in\{0,1\}$,
defined by the following relation and the multiplication rule
\begin{eqnarray}
  \oo{\alpha}_x (\underline{s}) = \delta_{\alpha,s_x},\qquad
  (\oo{\alpha}_x \oo{\beta}_y)(\underline{s}) =
  \oo{\alpha}_x (\underline{s})\,
  \oo{\beta}_y (\underline{s}),
\end{eqnarray}
where $\underline{s}=(\ldots,s_{-1},s_0,s_1,\ldots)$ is an arbitrary configuration of occupied and empty sites.
A product of  local observables on $r$ consecutive sites centered around $x$ is denoted by
\begin{eqnarray}
\label{locob}
  \oo{\alpha_1\alpha_2\ldots\alpha_r}_x \equiv
  \oo{\alpha_1}_{x-\lfloor\frac{r}{2}\rfloor}
  \oo{\alpha_2}_{x-\lfloor\frac{r}{2}\rfloor+1}
  \cdots
  \oo{\alpha_r}_{x+\lfloor\frac{r-1}{2}\rfloor},
\end{eqnarray}
and spans a complete function algebra over a finite sublatice ${\cal A}_{[y,z]}=\bigotimes_{x=y}^z {\cal A}_x$, with $y=x-\lfloor\frac{r}{2}\rfloor$, $z=x+\lfloor\frac{r-1}{2}\rfloor$.
For conciseness we sometimes omit the subscript, in which case it is assumed to be $0$,
\begin{eqnarray}
  \oo{\alpha_1\alpha_2\ldots\alpha_r} \equiv
  \oo{\alpha_1\alpha_2\ldots\alpha_r}_0.
\end{eqnarray}
The quasilocal algebra ${\cal A}$ can then be understood as an appropriate norm-closure of an inclusive sequence ${\cal A}_{[-z,z]} \subset {\cal A}_{[-z-1,z+1]} \subset  {\cal A}_{[-z-2,z+2]} \cdots$.

Let us introduce an \emph{identity observable} $\one_x=\oo{0}_x+\oo{1}_x$, a unit element in ${\cal A}_x$. Any observable $a \in {\cal A}$ is preserved
under multiplication by $\one_x$ (which in fact represent the same (unit) element in ${\cal A}$, $\one_x\equiv \one$)
\begin{eqnarray}
  a \one_x = \one_x  a = a,
\end{eqnarray}
therefore we can extend the support of any local observable multiplying by any number
of identities at the edges, e.g.\
\begin{eqnarray}\fl
  \oo{\alpha_1\alpha_2\ldots\alpha_r}_x &\equiv
  \one_{x-\lfloor\frac{r+2}{2}\rfloor}
  \cdot
  \oo{\alpha_1\alpha_2\ldots\alpha_r}_x
  \cdot
  \one_{x+\lfloor\frac{r+1}{2}\rfloor}\equiv\\
  \fl \nonumber
  &\equiv
  \oo{0\alpha_1\alpha_2\ldots\alpha_r0}_x
  +\oo{0\alpha_1\alpha_2\ldots\alpha_r1}_x
  +\oo{1\alpha_1\alpha_2\ldots\alpha_r0}_x
  +\oo{1\alpha_1\alpha_2\ldots\alpha_r1}_x.
\end{eqnarray}

\subsection{Expectation values and states}

For discussing statistical mechanics of the RCA 54, we introduce the notion of separable states, for which $ \{s_x\}_{x\in \ZZ} $ are Bernoulli random variables, corresponding to independent probability distributions 
$ p_x : s_x \to [0,1]$ for all $ x\in \ZZ $. 
The expectation values of a local observable
$\oo{\alpha_1\alpha_2\ldots\alpha_r}_x$ in a separable state $ p $, $ p\equiv (\dots,\{p_{-1}(0),p_{-1}(1)\},\{p_0(0),p_0(1)\},\{p_1(0),p_1(1)\} \dots) $, is given by the prescription
\begin{eqnarray}
  \langle \oo{\alpha_1\alpha_2\ldots\alpha_r}_x \rangle_p=p_{{x-\lfloor\frac{r}{2}\rfloor}}(\alpha_1)\cdot
  p_{{x-\lfloor\frac{r}{2}\rfloor+1}}(\alpha_2)
  \cdots
  p_{{x+\lfloor\frac{r-1}{2}\rfloor}}(\alpha_r).
\end{eqnarray}
For example, in the following sections we will consider two particular initial states:
\begin{enumerate}
	\item A \emph{maximum entropy state} in section \ref{sec:Cxt} defined by
	\begin{eqnarray}
	\label{inft}
	p_x(0)=p_x(1)=1/2, \quad \forall x \in \mathbb{Z}.
	\end{eqnarray}
	\item An \emph{inhomogeneous initial state} in section \ref{sec:inhQ} defined by
	\begin{eqnarray}
	\label{ihq}
	\begin{cases}
	p_x(0)=p_x(1)=1/2, \,\;\;\quad \mbox{ for } x\leq 0 \\
	p_x(0)=1, \quad p_x(1)=0. \quad \mbox{ for } x> 0
	\end{cases}
	\end{eqnarray}
\end{enumerate}
One could as well consider expectation values of local observables in a more general setup, where the states are not necessarily separable but satisfy a general clustering property, i.e. the
distribution of variable $s_x$ may depend on the values $\{s_{y}\}$ for $y$ sufficiently close to $x$, i.e., for $|x-y|\to \infty$, $\ave{[\alpha]_x [\beta]_y}_p \to \ave{[\alpha]_x}_p \ave{[\alpha]_y}_p$.

\subsection{Time evolution of local observables}
The dynamics on the configuration space induces the time-evolution of local observables, which corresponds to the time automorphism of the quasi-local algebra $ {\cal A} $. For any
observable $a \in \mathcal{A}$, we define its 
time evolved version $a^t \in \mathcal{A}$ as
\begin{eqnarray}
a^t(\underline{s}^0)=a(\underline{s}^t).
\end{eqnarray} 
Explicitly, a local time automorphism $U_x$ is defined by its action on 3-site local observables by
\begin{eqnarray}
  U_x\oo{\alpha\ \beta\ \gamma}_y=\begin{cases}
    \oo{\alpha\ \chi(\alpha,\beta,\gamma)\ \gamma}_y;& x=y,\\
    \oo{\alpha\ \beta\ \gamma}_y;& |x-y|\ge 2,
  \end{cases}
\end{eqnarray}
and is extended to any quasilocal observable $a\in {\cal A}$ by linearity and continuity.
The complete time evolution of the observables
\begin{eqnarray}
a^{t+1}=U(t)a^t,
\end{eqnarray}
is then given by composed linear automorphism $ U(t) $, which depends on the parity of time $ t $
\begin{eqnarray}
U(t)=\begin{cases}
\prod_{x\in 2\mathbb{Z}}U_x;& t\equiv 0\pmod 2,\\
\prod_{x\in 2\mathbb{Z}+1}U_x;& t\equiv 1\pmod 2.
\end{cases}
\end{eqnarray}

\section{Construction of time-dependent matrix product ansatz}\label{sec:tMPA}
In this section we present our main result, which is the derivation of the complete dynamics of local observables in terms of tMPA.
Using the homomorphism property of the time automorphism
\begin{eqnarray}
  \oo{\alpha_1\alpha_2\ldots\alpha_r}^t_x \equiv
  \oo{\alpha_1}_{x-\lfloor\frac{r}{2}\rfloor}^t
  \oo{\alpha_2}_{x-\lfloor\frac{r}{2}\rfloor+1}^t
  \cdots
  \oo{\alpha_r}_{x+\lfloor\frac{r-1}{2}\rfloor}^t,
\end{eqnarray}
it is possible to construct the tMPA of any local observable by tensor multiplying tMPAs for the single site observables.
Additionally, the stationarity of the identity observable $\one^t_x\equiv \one$
implies that the time evolution of the observable $\oo{0}_x$ can be expressed
in terms of the observable $\oo{1}_x$ as 
\begin{eqnarray}
\oo{0}_x^t=\one-\oo{1}_x^t.
\end{eqnarray}
To construct the tMPA of an arbitrary local observable it is thus sufficient to consider the time propagation of $ \oo{1}_x $.
The problem can be further reduced by noting that the time propagation of the shifted observable can be obtained by appropriately shifting the time propagated observable centered at the origin
\begin{eqnarray}
  \oo{1}_x^t =
  \begin{cases}
    \eta_x \left( \oo{1}^t\right);& x \equiv 0\pmod{2},\\
    \eta_x \left( \oo{1}^{t-1}\right);& x \equiv 1\pmod{2}.
  \end{cases}
\end{eqnarray}
Here $\eta_x$ is the lattice shift automorphism of ${\cal A}$ defined as $\eta_x(\oo{\underline{s}}_y)\equiv\oo{\underline{s}}_{x+y}$.

\begin{thm}\label{theorem}
The time evolution of the local observable $ [1] $ reads
\begin{eqnarray}
\oo{1}^t =\smashoperator{\sum_{s_{-t},\ldots,s_t \in \{0,1\}}} c_{s_{-t},\ldots,s_t} (t)
\oo{s_{-t} s_{-t+1} \cdots s_{t}},
\end{eqnarray}
where the amplitudes $ c_{s_{-t},\ldots,s_t} (t) $ can be represented in terms of the tMPA
\begin{eqnarray}
c_{s_{-t},\ldots s_t}(t)&=&\bra{l(t)}V_{s_{-t}} W_{s_{-t+1}}V_{s_{-t+2}}\cdots W_{s_{t-1}} V_{s_t}\ket{r} \nonumber \\
&+&  \bra{l^{\prime}}V^{\prime}_{s_{-t}} W^{\prime}_{s_{-t+1}} V^{\prime}_{s_{-t+2}}\cdots W^{\prime}_{s_{t-1}}V^{\prime}_{s_t}\ket{r^{\prime}(t)}.
\end{eqnarray}
$ V_{s},\ W_{s},\ V'_{s},\ W'_{s}\in  \text{End}(\mathcal{V}) $, $s\in\{0,1\}$, are linear operators over an infinite dimensional auxiliary Hilbert space $ \mathcal{V}=\lsp\{|c,w,n,a\rangle;\ c,w\in \mathbb{N}_0,\ n\in\{0,1,2\},\ a\in\{0,1\}\}$, 
and can be explicitly expressed in terms of ladder operators
\begin{eqnarray}\label{eq:DefLadderOperators}
\eqalign{
	\mathbf{c}^{+}= \smashoperator{\sum_{c,w,n,a}} \ket{c+1,w,n,a}\bra{c,w,n,a},
	\qquad &\mathbf{c}^{-}=\left(\mathbf{c}^{+}\right)^{T},\\
	\mathbf{w}^{+}=\smashoperator{\sum_{c,w,n,a}} \ket{c,w+1,n,a}\bra{c,w,n,a},
	\qquad &\mathbf{w}^{-}=\left(\mathbf{w}^{+}\right)^{T},
}
\end{eqnarray}
and projectors
\begin{eqnarray}
  \eqalign{
  	\mathbf{e}_{c_2 w_2 n_2 a_2, c_1 w_1 n_1 a_1}=\ket{c_2,w_2,n_2,a_2}\bra{c_1,w_1,n_1,a_1},\\
  	\mathbf{e}_{n_2 a_2, n_1 a_1}=\sum_{c,w} \ket{c,w,n_2,a_2}\bra{c,w,n_1,a_1},
  }
\end{eqnarray}
as
\begin{eqnarray}\label{eq:matricesWV}
\fl
\eqalign{
	V_0 &= \mathbf{e}_{00,00}+\mathbf{e}_{10,00}+\mathbf{e}_{20,00}
	+ \mathbf{c}^{+} \mathbf{e}_{10,01}
	+ \mathbf{e}_{01,01}
	+\mathbf{c}^{+}\mathbf{w}^{+} \mathbf{e}_{11,01}
	+\mathbf{e}_{21,01}+\\
	&+ \mathbf{e}_{0001,0001}
	+ \mathbf{e}_{0011,0001}
	+ \mathbf{e}_{0021,0001},\\
	V_1 &=
	\mathbf{e}_{00,10}+\mathbf{e}_{10,20}+\mathbf{e}_{20,20}
    +\mathbf{e}_{00,11}+\mathbf{e}_{10,21}+\mathbf{e}_{20,21}
	+\mathbf{e}_{01,11}
	+\mathbf{w}^{+} \mathbf{e}_{11,21} +\mathbf{w}^{+}\mathbf{e}_{21,21}
	+\\&+ \mathbf{e}_{0001,0011}
	+ \mathbf{e}_{0011,0021}
	+ \mathbf{e}_{0021,0021},\\
	W_0 &=
	\mathbf{c}^{-} \mathbf{w}^{+} \left(\mathbf{e}_{00,00}+\mathbf{e}_{10,00}+\mathbf{e}_{20,00}\right)
	+ \mathbf{w}^{+} \mathbf{e}_{10,01}
	+\mathbf{w}^{+} \mathbf{e}_{01,01}
	+\mathbf{c}^{+}\left(\mathbf{w}^{+}\right)^2 \mathbf{e}_{11,01}
	+\\&+\mathbf{w}^{+} \mathbf{e}_{21,01}
	+ \mathbf{e}_{1111,0001}
	+ \mathbf{e}_{0001,0001}
	+ \mathbf{e}_{0011,0001}
	+ \mathbf{e}_{0021,0001},\\
	W_1 &=
	\mathbf{c}^{-}\mathbf{w}^{+}\left(\mathbf{e}_{00,10}+\mathbf{e}_{10,20}+\mathbf{e}_{20,20}\right)
	+\mathbf{w}^{+}\mathbf{e}_{01,11}
	+\mathbf{c}^{+}\mathbf{w}^{+}\mathbf{e}_{11,21}
	+\mathbf{c}^{+}\mathbf{w}^{+}\mathbf{e}_{21,21}
	+\\&+ \mathbf{e}_{0001,0011}
	+ \mathbf{e}_{0011,0021}
	+ \mathbf{e}_{0021,0021},
}
\end{eqnarray}
and
\begin{eqnarray}\label{eq:matricesWVprime}
\eqalign{
	V_0^{\prime}=V_0^T-\left(\mathbf{e}_{0001,1111}+\mathbf{e}_{0101,1211}+\mathbf{e}_{0101,1110}\right),\\
	V_1^{\prime}=V_1^T,\\
	W_0^{\prime}=W_0^T-\left(\mathbf{e}_{0001,1111}+\mathbf{e}_{0000,1211}\right),\\
	W_1^{\prime}=W_1^T-\left(\mathbf{e}_{0021,1111}+\mathbf{e}_{0021,1121}+\mathbf{e}_{0121,1211}+\mathbf{e}_{0121,1221}\right).
}
\end{eqnarray}
The time-dependent auxiliary space boundary vectors take the following form
\begin{eqnarray}
\eqalign{
	\bra{l(t)}=\bra{0,t,0,0},\\
	\ket{r}=\ket{0,0,0,0}+\ket{0,0,0,1}+\ket{0,0,0,2},\\
	\bra{l^{\prime}}=\bra{0,0,0,1}+\bra{0,0,1,1}+\bra{0,0,2,1}+\bra{0,1,0,1}+\bra{0,1,2,1},\\
	\ket{r^{\prime}(t)}=\ket{0,t+1,0,0}.
}
\end{eqnarray}
\end{thm}

\begin{rem}
  Before dwelling into the full-blown proof, let us first
elucidate the main idea leading to a compact description of the dynamics
 and its properties.

First of all note that the reduction of the dynamics of the ultra-local
observable to the sub-lattice of the size $ 2t+1 $ around the origin is a
consequence of the locality of the time evolution operator $ U $, which maps
the subalgebra of local observables ${\cal A}_{[-t,t]}$ to the subalgebra of
local observables with the increased support ${\cal A}_{[-t-1,t+1]}$, and can
thus be equivalently expressed in terms of the reduced propagator
\begin{eqnarray}
  U_{[-t-1,t+1]}\equiv U_{-t}\,U_{-t+2} \cdots U_{t}.
\end{eqnarray}
To be more concise, the dynamics of the local observable $\oo{s_{-t}\cdots s_t}$
is completely determined by the following prescription
\begin{eqnarray}
  &U(t) \oo{s_{-t}\cdots s_t} \equiv\\
  &U_{[-t-1,t+1]}(\oo{0s_{-t}\cdots s_t0}
  +\oo{0s_{-t}\cdots s_t1}
  +\oo{1s_{-t}\cdots s_t0}
  +\oo{1s_{-t}\cdots s_t1}). \nonumber
\end{eqnarray}

The fact that we have a deterministic dynamics has two immediate consequences.
Firstly, the coefficients $c_{s_{-t},\ldots s_t}(t)$ can only be $0$, corresponding to the \emph{inaccessible} configurations, or $1$, corresponding to the \emph{accessible} configurations. Secondly, the number of accessible configurations at time $t$
is $4^{t}$ which is half of all possible distinct configurations $2^{2t+1}$.

The construction of the tMPA relies on an explicit identification of all
accessible configurations.
The initial configuration is $\oo{1}$, describing all possible states with at
least one soliton traversing through the origin at time $ t=0 $. At time $t$,
the soliton originating from the center resides between the lattice sites
$x=-t$ and $x=t$, i.e.\ the section of the lattice referred to
as a \emph{light-cone}. The exact position of the soliton is determined by the number
of scatterings $c$. An example is shown in the Figure~\ref{fig:OneChosenSoliton}.
\begin{figure}
  \centering
  \reflectbox{\includegraphics[width=\linewidth]{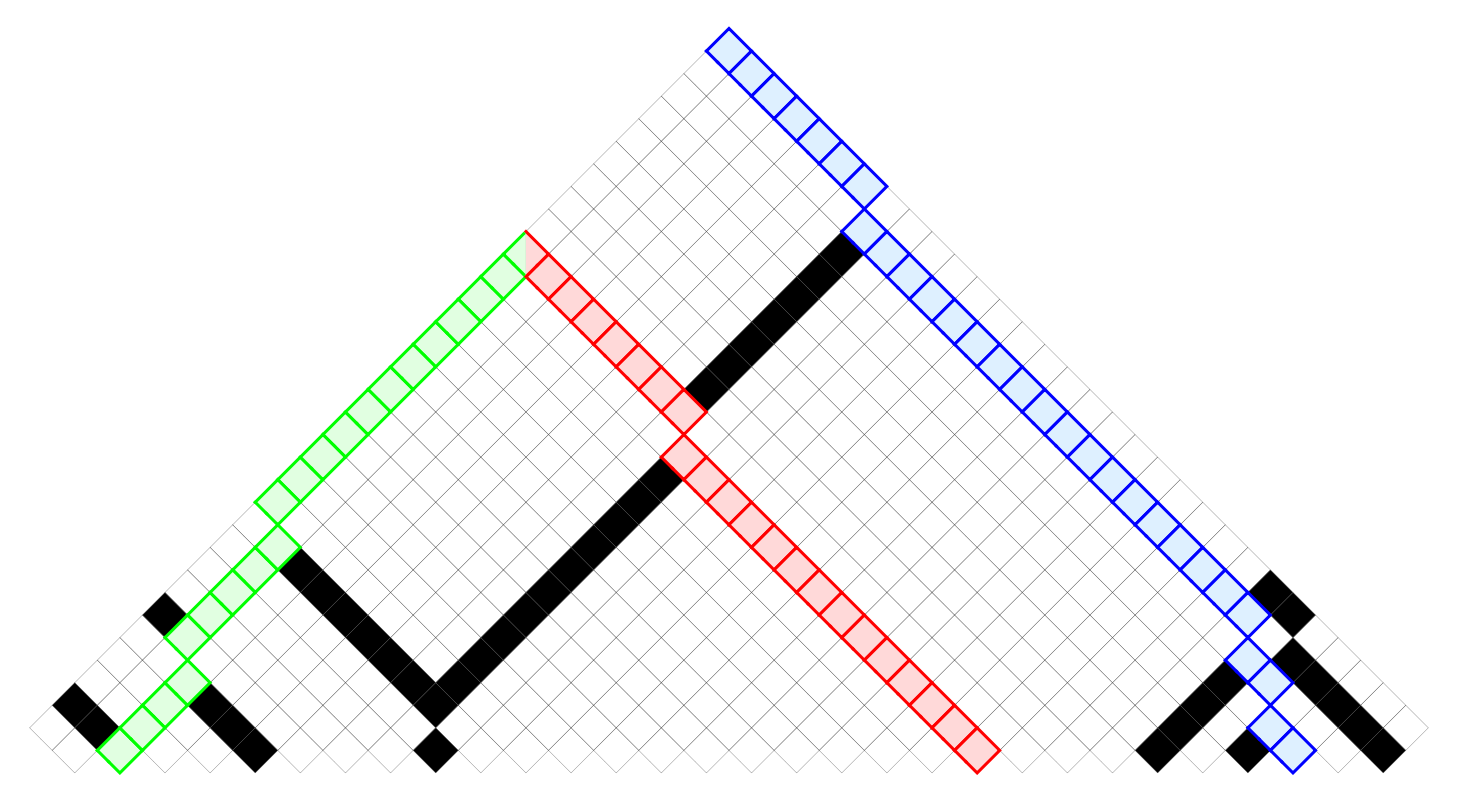}}
  \caption{\label{fig:OneChosenSoliton} An example of an allowed configuration
    (the bottom-most saw). The red, blue and green-colored sites denote three
    distinct solitons. The blue soliton goes through
    the site $(0,0)$, while the red and green solitons originate outside of the light-cone.
    Alternatively, we can think of this as solitons that start at the bottom
    and propagate in the negative time.
    The blue soliton  goes through the top site, while the red and green one
    escape the light-cone and cannot reach the origin.}
\end{figure}

The configuration at time $t$ contains the complete information about the
particle content and the scatterings, implying that we can propagate any
configuration backwards in time in order to determine whether one of the
solitons
originated from the central position. The tMPA is constructed so that it traces
every particle backwards in time, by counting the number of scatterings, and
determines whether a given particle ends up at the origin. If this is the case,
tMPA coefficient yields $ 1 $, otherwise the contribution vanishes.
\end{rem}

\begin{proof}
The proof of theorem 1 consists of two parts. In the first part we derive the tMPA for the states in which the soliton emerging from the center is a left mover, and in the second part for the central right movers.
  \subsection*{tMPA for the left mover emerging from the center}
  Let us consider a configuration $\oo{s_{-t}\cdots s_t} \in \{0,1\}^{2t+1}$. Using the tMPA we
  scan a given configuration site by site, starting from the left edge of the
  light-cone and moving towards the right edge. Whenever we encounter a left
  mover, which we dub the probe, we count the number of solitons on its right
  in order to determine, whether it originated from the center. To encode the
  soliton counting procedure, we introduce four auxiliary degrees of freedom,
  $\ket{c,w,n,a}$:
\begin{enumerate}
  \item \emph{The activation bit}, $a\in\{0,1\}$, tells us whether we are on the
    left or the right side of the probe. If the activation bit is turned off, i.e. $a=0$, the state splits into two parts whenever we encounter a left mover. The first part corresponds to the value $ a=0 $, describing the situation in which the left mover is not a probe, while the second part represents the opposite case, with $ a=1 $. In any other situation the activation bit remains unchanged.
  \item \emph{The collision counter}  $c\in\mathbb{N}_0$, represents the number
    of scatterings that the probe has to undergo in order to reach
    the origin in the inverse scattering procedure, and at the same time
    it distinguishes between the different probes of the same state. While $a=0$,
    the collision counter increases by $1$ every two sites. If $a=1$, the collision counter
    decreases by $1$ whenever a left moving soliton that scattered with the probe is encountered.
    If at the right edge of the light cone the collision counter is zero,
    the probe passed through the origin.
  \item \emph{The scattering width}  $w\in\mathbb{N}_0$, keeps
    track of the scatterings of the right movers after the probe.
    At the left edge the width is equal to time $t$, and after every two sites it decreases by $1$.
    Additionally, the width changes as $w\to w-1$, whenever a left mover on the right side of the probe is encountered. If the width drops to $0$, the right movers that we meet did not scatter
    with the probe.
  \item \emph{The occupation counter} $n\in\{0,1,2\}$ provides additional
    information about the particle content needed to appropriately change $w$ and $c$. Explicitly,
    $n=0$ if the current site is empty, $n=1$ if the site is full and the left neighbor is empty,
    and $n=2$ if the site and the left neighbor are both occupied.
\end{enumerate}
In the initial state the collision counter $ c $ is $0$, and the width $w$ is set to the number of time-slices $ t $,
\begin{eqnarray}
  \bra{l(t)}=\bra{0,t,0,0}.
\end{eqnarray}
The right boundary vector has nonzero overlap with vectors that correspond
to a probe that passed through the origin at time $ t=0 $, i.e.\ $c=w=0$ and $a=1$, while the occupation
number can be arbitrary,
\begin{eqnarray}
  \ket{r}=\ket{0,0,0,1}+\ket{0,0,1,1}+\ket{0,0,2,1}.
\end{eqnarray}
Before constructing the matrix elements, let us first introduce 
the projector to the subspace with $a=0$,
\begin{eqnarray}
  P_0=\smashoperator{\sum_{c,w,n}}\ket{c,w,n,0}\bra{c,w,n,0}=P_0^{T}=P_0^2,
\end{eqnarray}
and consider three different regimes.

\subsubsection*{Left side of the probe}
Before choosing the probe,
the width and the required number of scatterings have to be adjusted,
therefore the left action of the restricted matrices $W_s P_0$, $V_s P_0$ corresponds to
\begin{eqnarray}
  \eqalign{
    \bra{c,w,n,0} V_s P_0 = \bra{c,w,s\cdot\min\{n+1,2\},0},\\
    \bra{c,w,n,0} W_s P_0 = \bra{c+1,w-1,s\cdot\min\{n+1,2\},0}.
  }
\end{eqnarray}

\subsubsection*{Choosing the probe}
Whenever a left moving soliton is encountered, an additional vector with $a=1$ is created.
There are $4$ such configurations. The two simpler ones correspond to a soliton
that appears on the right diagonal,
\begin{eqnarray}\label{eq:RightMovingCreationSimple}
  \begin{tikzpicture}
    \emptyrectangle{0}{1};
    \fullrectangle{1}{0}
    \fullrectangle{2}{1};
  \end{tikzpicture}\qquad
  \begin{tikzpicture}
    \fullrectangle{0}{1};
    \fullrectangle{1}{0}
    \fullrectangle{2}{1};
  \end{tikzpicture},
\end{eqnarray}
which implies
\begin{eqnarray}
  \bra{c,w,1,0}V_1 (1-P_0) = \bra{c,w,2,0}V_1 (1-P_0) = \bra{c,w,2,1}.
\end{eqnarray}
The other two configurations correspond to encountering a soliton while it is scattering,
\begin{eqnarray}\label{eq:RightMovingCreationComplicated}
  \eqalign{
    \begin{tikzpicture}
      \emptyrectangle{0}{0};
      \fullrectangle{1}{1}
      \halfrectangle{2}{2};
      \emptyrectangle{2}{0};
    \end{tikzpicture}\qquad 
    \begin{tikzpicture}
      \emptyrectangle{0}{1};
      \fullrectangle{1}{0}
      \halfrectangle{1}{2};
      \halfrectangle{2}{3};
      \emptyrectangle{2}{1};
    \end{tikzpicture},
  }
\end{eqnarray}
where the grey colored squares represent the soliton's estimated path in absence of any
additional encounters. The appropriate matrix elements are the following,
\begin{eqnarray}
  \eqalign{
    \bra{c,w,1,0}W_0(1-P_0)=\bra{c,w-1,0,1},\\
    \bra{c,w,1,0}V_0(1-P_0)=\bra{c-1,w,0,1}.
  }
\end{eqnarray}

\subsubsection*{Right side of the probe}
Let us assume that $w>0$. Once a vector with $a=1$ is produced, the collision counter has to be
decreased whenever a right mover is encountered,
while the width $w$ decreases every two sites and additionally whenever a left mover is met. Explicitly, there are two possible configurations of a right mover appearing,
\begin{eqnarray}
\eqalign{
	\begin{tikzpicture}
	\emptyrectangle{0}{0};
	\fullrectangle{1}{1};
	\fullrectangle{2}{0};
	\end{tikzpicture}\qquad 
	\begin{tikzpicture}
	\fullrectangle{0}{0};
	\fullrectangle{1}{1};
	\fullrectangle{2}{0};
	\end{tikzpicture},
}
\end{eqnarray}
which are described by the following matrix elements,
\begin{eqnarray}
\bra{c,w,1,1}W_1 = \bra{c,w,2,1} W_1 = \bra{c-1,w-1,2,1}.
\end{eqnarray}
The two configurations from~\eqref{eq:RightMovingCreationSimple}
correspond to a process of encountering a left moving soliton,
which is described by the following two matrix elements
\begin{eqnarray}
\bra{c,w,1,1}V_1 = \bra{c,w,2,1} V_1  = \bra{c,w-1,2,1}.
\end{eqnarray}
If the scattering solitons are encountered (eq.\ \eqref{eq:RightMovingCreationComplicated}),
we have to decrease the width and the collision counter at the same time,
which amounts to the following,
\begin{eqnarray}
\eqalign{
	\bra{c,w,1,1}W_0 =\begin{cases}
	0; & w=1\text{ and }c\neq1,\\
	\bra{0,0,0,1};& w=c=1,\\
	\bra{c-1,w-2,0,1}; & w\geq 2.
	\end{cases}\\
	\bra{c,w,1,1}V_0 = \bra{c-1,w-1,0,1}.
}
\end{eqnarray}
In all the remaining cases,
\begin{eqnarray}\fl
  \eqalign{
    \begin{tikzpicture}
      \unknownrectangle{0}{0};
      \emptyrectangle{1}{1};
      \emptyrectangle{2}{0};
    \end{tikzpicture}\qquad 
    \begin{tikzpicture}
      \unknownrectangle{0}{0};
      \emptyrectangle{1}{1};
      \fullrectangle{2}{0};
    \end{tikzpicture}\qquad 
    \begin{tikzpicture}
      \fullrectangle{0}{0};
      \fullrectangle{1}{1}
      \emptyrectangle{2}{0};
    \end{tikzpicture}\qquad
    \begin{tikzpicture}
      \unknownrectangle{0}{1};
      \emptyrectangle{1}{0}
      \emptyrectangle{2}{1};
    \end{tikzpicture}\qquad
    \begin{tikzpicture}
      \unknownrectangle{0}{1};
      \emptyrectangle{1}{0}
      \fullrectangle{2}{1};
    \end{tikzpicture}\qquad
    \begin{tikzpicture}
      \fullrectangle{0}{1};
      \fullrectangle{1}{0}
      \emptyrectangle{2}{1};
    \end{tikzpicture},\\
  }
\end{eqnarray}
there are no additional decreases of $w$ and $c$, therefore
\begin{eqnarray}
  \eqalign{
    \bra{c,w,0,1}W_0=\bra{c,w-1,0,1},\qquad
    &\bra{c,w,0,1}W_1=\bra{c,w-1,1,1},\\
    \bra{c,w,2,1}W_0=\bra{c,w-1,0,1},
    &\bra{c,w,0,1}V_0=\bra{c,w,0,1},\\
    \bra{c,w,0,1}V_1=\bra{c,w,1,1},
    &\bra{c,w,2,1}V_0=\bra{c,w,0,1}.
  }
\end{eqnarray}
The right movers that are encountered after the width $ w $ drops to $ 0 $ did
not scatter with the probe, therefore $c$ should
not decrease anymore. The probe reached the origin only if the value of the collision counter $ c $ is $ 0 $ on the right side of the light-cone, inducing the following matrix elements
\begin{eqnarray}\label{eq:subspaceweq0}\fl
  \bra{c,0,n,1}W_s = \bra{c,0,n,1} V_s=\begin{cases}
    \bra{0,0,s\cdot \max\{2,n+1\},1};& c=0,\\
    0;& c>0,
  \end{cases}
\end{eqnarray}
which completes the construction of the tMPA for the left movers~\eqref{eq:matricesWV}.

\subsection*{tMPA for the central right movers}
The tMPA of the right movers can be derived in a similar fashion, by reversing the direction of all solitons. This corresponds to exchanging the roles of the left and the right boundary vectors, and transposing the auxiliary matrices $W_s$ and $V_s$.  However, we have to additionally exclude all of
the configurations that were captured by the tMPA for the left movers,
i.e.\ the configurations where both the left and the right mover are emitted from the origin.
Up to time $t=2$, the configurations that should be excluded are
\begin{eqnarray}\label{eq:ForbiddenConfigs}
  \eqalign{
    \begin{tikzpicture}
      \fullrectangle{0}{0};
      \fullrectangle{-1}{-1};
      \fullrectangle{1}{-1};
      \fullrectangle{-2}{-2};
      \emptyrectangle{0}{-2};
      \emptyrectangle{2}{-2};
      \emptyrectangle{-1}{-3};
      \fullrectangle{1}{-3};
    \end{tikzpicture}\qquad 
    \begin{tikzpicture}
      \fullrectangle{0}{0};
      \fullrectangle{-1}{-1};
      \fullrectangle{1}{-1};
      \fullrectangle{-2}{-2};
      \emptyrectangle{0}{-2};
      \fullrectangle{2}{-2};
      \emptyrectangle{-1}{-3};
      \emptyrectangle{1}{-3};
    \end{tikzpicture}\qquad 
    \begin{tikzpicture}
      \fullrectangle{0}{0};
      \fullrectangle{-1}{-1};
      \fullrectangle{1}{-1};
      \emptyrectangle{-2}{-2};
      \emptyrectangle{0}{-2};
      \emptyrectangle{2}{-2};
      \fullrectangle{-1}{-3};
      \fullrectangle{1}{-3};
    \end{tikzpicture}\qquad 
    \begin{tikzpicture}
      \fullrectangle{0}{0};
      \fullrectangle{-1}{-1};
      \fullrectangle{1}{-1};
      \emptyrectangle{-2}{-2};
      \emptyrectangle{0}{-2};
      \fullrectangle{2}{-2};
      \fullrectangle{-1}{-3};
      \emptyrectangle{1}{-3};
    \end{tikzpicture}\\
    \begin{tikzpicture}
      \fullrectangle{0}{0};
      \emptyrectangle{-1}{-1};
      \emptyrectangle{1}{-1};
      \emptyrectangle{-2}{-2};
      \fullrectangle{0}{-2};
      \fullrectangle{2}{-2};
      \fullrectangle{-1}{-3};
      \fullrectangle{1}{-3};
    \end{tikzpicture}\qquad
    \begin{tikzpicture}
      \fullrectangle{0}{0};
      \emptyrectangle{-1}{-1};
      \emptyrectangle{1}{-1};
      \emptyrectangle{-2}{-2};
      \fullrectangle{0}{-2};
      \emptyrectangle{2}{-2};
      \fullrectangle{-1}{-3};
      \fullrectangle{1}{-3};
    \end{tikzpicture}\qquad
    \begin{tikzpicture}
      \fullrectangle{0}{0};
      \emptyrectangle{-1}{-1};
      \emptyrectangle{1}{-1};
      \fullrectangle{-2}{-2};
      \fullrectangle{0}{-2};
      \fullrectangle{2}{-2};
      \fullrectangle{-1}{-3};
      \fullrectangle{1}{-3};
    \end{tikzpicture}\qquad
    \begin{tikzpicture}
      \fullrectangle{0}{0};
      \emptyrectangle{-1}{-1};
      \emptyrectangle{1}{-1};
      \fullrectangle{-2}{-2};
      \fullrectangle{0}{-2};
      \emptyrectangle{2}{-2};
      \fullrectangle{-1}{-3};
      \fullrectangle{1}{-3};
    \end{tikzpicture}
  }.
\end{eqnarray}
This can be achieved  by considering the alternative boundary vectors
\begin{eqnarray}
  \eqalign{
    \ket{r^{\prime}(t)}=\ket{0,t+1,0,0},\\
    \bra{l^{\prime}}=\bra{0,0,0,1}+\bra{0,0,1,1}+\bra{0,0,2,1}+\bra{0,1,0,1}+\bra{0,1,2,1},
  }
\end{eqnarray}
and by changing the tMPA matrices, so that the following holds
\begin{eqnarray}
  \eqalign{
    W_0^{\prime}\ket{1,1,1,1}=W_0^{\prime}\ket{1,2,1,1}=0,\\
    W_1^{\prime}\ket{1,1,1,1}=W_1^{\prime}\ket{1,2,1,1}=0,\\
    W_1^{\prime}\ket{1,1,2,1}=W_1^{\prime}\ket{1,2,2,1}=0,\\
    V_0^{\prime}\ket{1,1,1,1}=V_0^{\prime}\ket{1,2,1,1}=0,\\
    (1-P_0)V_0^{\prime}\ket{1,1,1,0}=0.
  }
\end{eqnarray}
\end{proof}

\section{Time-dependent density profile after inhomogeneous quench}\label{sec:inhQ}
In this and the following section we will consider two physically relevant applications of the tMPA. The first example is an explicit calculation of the particle density profile following the inhomogeneous
quench. 
The density profile corresponds to the probability of observing a particle at site $x$ and time $t$
\begin{eqnarray}
\hat{\rho}(x,t)=\ave{\oo{1}_x}_{p^t}=\ave{\oo{1}_x^{-t}}_p,
\end{eqnarray}
where the negative time propagation is given by
\begin{eqnarray}
\oo{1}_x^{-t} = \begin{cases}
\eta_x\left(\oo{1}^t\right);& x+t\equiv 1 \pmod{2},\\
\eta_x\left(\oo{1}^{t-1}\right);& x+t\equiv 0 \pmod{2}.
\end{cases}
\end{eqnarray}
At time $t=0$, the system is prepared in the state \eqref{ihq},
in which the probability of a site being occupied is $1/2$ on the left side of the chain, and $ 0 $ on the right side of the chain.

First of all note that the density profile changes only in the vicinity of the junction $-t\le x \le t$,
and can be efficiently expressed in terms of the density profile along the diagonal $ m $, $\rho(m,t)$,
\begin{eqnarray}
\hat{\rho}(x,t)=\begin{cases}
0; & t\le x,\\
\rho(\frac{t-x}{2},t-1);& x+t\equiv 0\!\pmod 2 \text{ and } -t< x < t,\\
\rho(\frac{t-x+1}{2},t);& x+t\equiv 1\pmod 2 \text{ and } -t< x < t,\\
\frac{1}{2};& x\le -t,
\end{cases}
\end{eqnarray}
 which can be calculated efficiently, using the tMPA representation
\begin{eqnarray}
  \eqalign{
    \rho(m,t)=2^{-2m}\left( L(m,t) + R(m,t)\right),\\
    L(m,t)=\bra{l(t)}\left((V_0+V_1)(W_0+W_1)\right)^m V_0 \left(W_0 V_0\right)^{t-m}\ket{r},\\
    R(m,t)=\bra{l^{\prime}}\left((V_0^{\prime}+V_1^{\prime})(W_0^{\prime}+W_1^{\prime})\right)^m 
    V_0^{\prime} \left(W_0^{\prime} V_0^{\prime}\right)^{t-m}\ket{r^{\prime}(t)}.
  }
\end{eqnarray}

In order to calculate the matrix elements $L(x,t)$ and $R(x,t)$, the following reduction can be employed. The action of any matrix  $M_s\in\{V_s^T,V_s^{\prime},W_s^{T},W_s^{\prime}\}$,
\begin{enumerate}
  \item on a vector with the state index $a=1$ does not produce a state from the subspace
  corresponding to $a=0$, i.e.\ $P_0 M_s (1-P_0)=0$,
  \item on a state with $ a=0 $ does not increase its collision ($c$) or its
    width ($w$) index,
  \item on the state $ \ket{0,0,n,1} $ produces a vector $ \ket{0,0,s\cdot \max\{n+1,2\}} $.
\end{enumerate}
\begin{figure}
  \centering
  \input{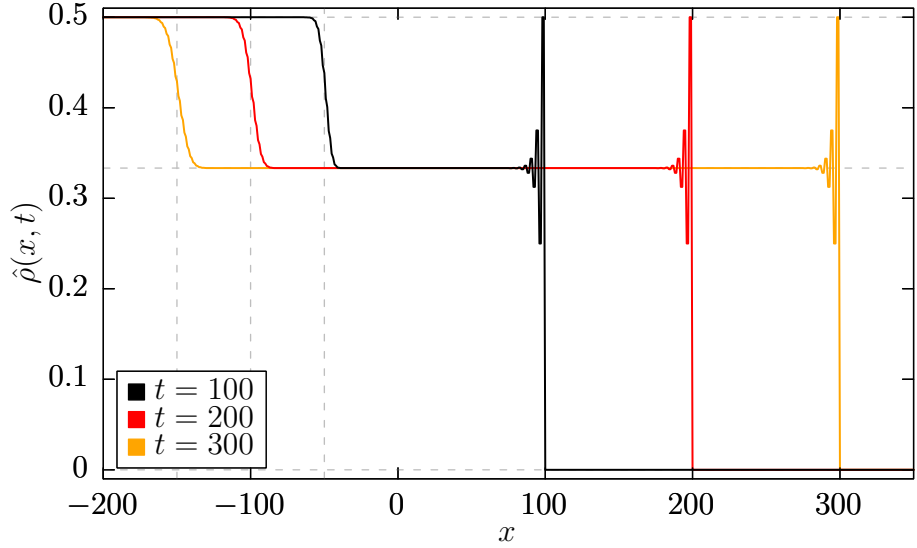}
  \caption{\label{fig:inhQplot}Density profile at different times $t$ after the quench.
    The ballistic front on the right moves with the velocity $1$ and its shape does not change.
    On the left side, the profile has a shape of an $\erf$ that moves with a velocity $1/2$
    and interpolates between $1/3$ towards the center and $1/2$ to the left. Its width scales
    as $\sim\sqrt{t}$.
  }
\end{figure}
Using these properties, we can calculate certain matrix elements explicitly,
see \ref{sec:actionOfT}. In particular the density profile reads (see
\ref{sec:appInhQ}),
\begin{eqnarray}\label{eq:shiftedRho}
  \eqalign{
    \rho(m,t)&=\frac{3}{8}\delta_{m,t}+\frac{1}{8}\left(\delta_{m,1}\delta_{t,1}+\delta_{m,2}\delta_{t,2}\right)+
    \frac{1}{4}\delta_{m,3}\delta_{t,4}+
    \\&+ \theta_{2m-t-3}2^{t-2m}\left(3\binom{2m-t-3}{t-m-1}+\binom{2m-t-3}{t-m-2}\right)+
    \\&+ \frac{1-\left(-\frac{1}{2}\right)^{m}}{3}
    -\frac{1}{2}\sum_{y=t-m}^{m-1}2^{-(m-1-y)} \binom{m-1-y}{y}+
    \\&+\frac{1}{8}\sum_{y=t-m}^{2m-t-3} 2^{-(2m-t-3)}\binom{2m-t-3}{y}
    +\frac{3}{16}\smashoperator[r]{\sum_{y=0}^{2m-t-4}} 2^{-y} \binom{y}{t-m-1},
  }
\end{eqnarray}
where $\theta_x$ is a discrete Heaviside function; $\theta_{x\ge 0}=1$ and
$\theta_{x<0}=0$. The profile is plotted for three distinct times in
Figure~\ref{fig:inhQplot}. From it, we can immediately identify two distinct regimes.

\subsection{Free regime}
The density profile is particularly simple in the region with the diagonal
index $m\le 2t/3$, where only a single term from the equation
\eqref{eq:shiftedRho} survives,
\begin{eqnarray}
  \rho\left(m\le \frac{2t}{3},t\right) = \frac{1-\left(-\frac{1}{2}\right)^m}{3}.
\end{eqnarray}
The density profile in this regime corresponds to the alternating exponential
decay centered around $1/3$, traveling with a maximal velocity $v_{\text{max}}=1$,
\begin{eqnarray}\label{eq:inhQprofileIntermediate}
  \hat{\rho}\left(t\ge x\ge-\frac{t}{3}+1,t\right)
  =\frac{1}{3}\left(1-\left(-\frac{1}{2}\right)^{\lceil\frac{t-x}{2}\rceil}\right).
\end{eqnarray}
\begin{figure}
  \centering
  \input{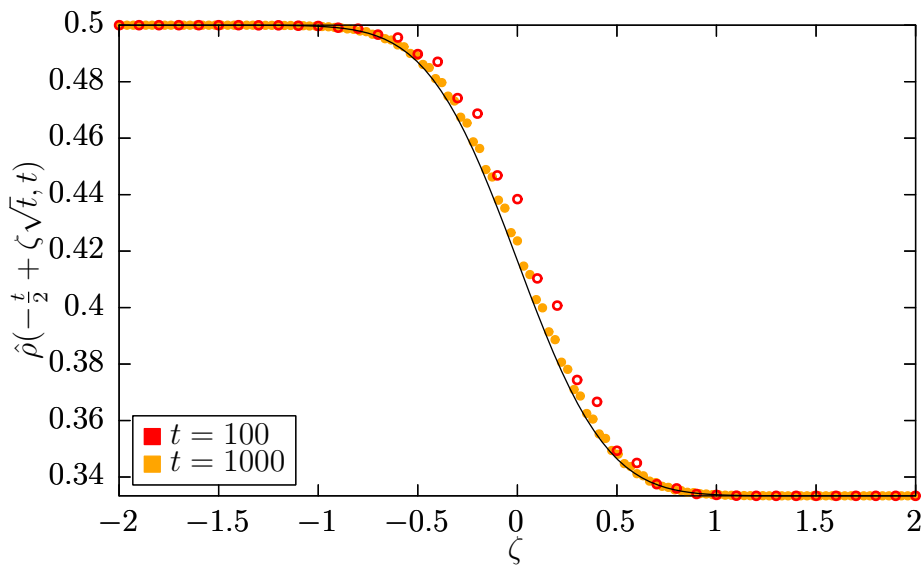}
  \caption{\label{fig:inhQasymptotics} The density profile $\hat{\rho}(x,t)$
    around $x=-\frac{t}{2}$. The solid curve denotes the asymptotic profile,
    $\hat{\rho}(\zeta)=\frac{1}{12}(5-\erf(2\zeta))$.
  }
\end{figure}
Note that the appearance of this regime is reminiscent of the generic situation occurring at low temperatures in any integrable model \cite{1742-5468-2018-3-033104}. For a more intuitive understanding of this regime see \ref{free_regime}.

\subsection{Thermalizing regime}
If $m>\frac{2t}{3}$ (and $t\ge 7$), the profile~\eqref{eq:shiftedRho}
can be expressed as 
\begin{eqnarray}\fl \label{eq:diffusiveShiftedProfile}
  \eqalign{
    \rho(m,t)&=\frac{3}{8}\delta_{m,t}+2^{t-2m-1}\binom{2m-t-3}{t-m-1}+
    \frac{1}{2}\sum_{y=0}^{t-m-1}2^{-(m-1-y)} \binom{m-1-y}{y}+
    \\&+\frac{1}{8}\sum_{y=t-m-2}^{2m-t-3} 2^{-(2m-t-3)}\binom{2m-t-3}{y}
    +\frac{3}{16}\smashoperator[r]{\sum_{y=0}^{2m-t-3}} 2^{-y} \binom{y}{t-m-1}.
  }
\end{eqnarray}
Asymptotically this reduces to
\begin{eqnarray}
  \lim_{t\to\infty}\rho\left(\frac{3t}{4}+\zeta \sqrt{t},t\right) = \frac{1}{12}\left(5+\erf(4\zeta)\right),
\end{eqnarray}
implying the following shape of the density profile
\begin{eqnarray}
  \hat{\rho}(\zeta)\equiv \lim_{t\to\infty}\hat{\rho}\left(-\frac{t}{2}+\zeta \sqrt{t},t\right)
  = \frac{1}{12}\left(5-\erf(2\zeta)\right).
\end{eqnarray}
The comparison of the profile at finite times and the asymptotic expression
is shown in Figure \ref{fig:inhQasymptotics}.

\section{Dynamic structure factor}\label{sec:Cxt}
In this section we obtain an explicit expression of the \emph{spatio-temporal density-density correlation function},
i.e.\ the real space-time expression for the dynamic structure factor,
\begin{eqnarray}
  C(x,t)=\langle\oo{1}_x \oo{1}^t\rangle_p-\langle\oo{1}_x\rangle_p\langle\oo{1}^t\rangle_p
  =\langle\oo{1}_x \oo{1}^t\rangle_p-\frac{1}{4},
\end{eqnarray}
where $ p $ is the maximum entropy state \eqref{inft}. The dynamic structure  factor corresponds to the probability that the particle, which is initially localized at the origin moves to the site $ x $ in time $ t $.

As a consequence of the staggered structure of the time evolution, the following holds
\begin{eqnarray}
  C(x,t)=\begin{cases}
    C(x,t-1); & x+t\equiv 0 \pmod{2},\\
    C(x,t+1); & x+t\equiv 1 \pmod{2},
  \end{cases}
\end{eqnarray}
implying that the generic expression for $C(x,t)$ can be obtained by considering only the cases with  $x+t\equiv 0 \pmod{2}$.
Under this assumption, the dynamic structure factor can be represented in terms of tMPA as
\begin{eqnarray}\fl
  C(x,t)=\frac{1}{2^{2t+1}}\left(\bra{l(t)} T^{\frac{x+t}{2}} V_1 \overline{T}^{t-{\frac{x+t}{2}}}\ket{r}+
    \bra{l^{\prime}} \overline{T}^{\prime\,{\frac{x+t}{2}}} V_1^{\prime}
  {T}^{\prime\,t-{\frac{x+t}{2}}}\ket{r^{\prime}(t)}\right)-\frac{1}{4},
\end{eqnarray}
with
\begin{eqnarray}
\eqalign{
	T^{\phantom{\prime}}=(V_0+V_1)(W_0+W_1),\qquad
	&\overline{T}^{\phantom{\prime}}=(W_0+W_1)(V_0+V_1),\\
	T^{\prime} = (W_0^{\prime}+W_1^{\prime})(V_0^{\prime}+V_1^{\prime}),
	&\overline{T}^{\prime} = (V_0^{\prime}+V_1^{\prime})(W_0^{\prime}+W_1^{\prime}).
}
\end{eqnarray}
In order to simplify the derivation of the structure factor, it proves useful to consider the rescaled difference $\Delta C(x,t)$, of two next-to-nearest neighboring correlation functions,
\begin{eqnarray}\fl
  \label{eq:exactDifferencesLR}
  \Delta C(x,t) &\equiv 2^{2t+1}\left(C(x+2,t)-C(x,t)\right)=\\ \nonumber \fl
  &= \underbrace{\bra{l(t)}T^{\frac{x+t}{2}}
    \left(T\,V_1-V_1\overline{T}\right)
  \overline{T}^{\frac{t-x}{2}-1}\ket{r}}_{\Delta C_l(\frac{x+t}{2},t)}
  -\underbrace{\bra{l^{\prime}}\overline{T}^{\prime \frac{x+t}{2}}
    \left(V_1^{\prime}T^{\prime}-\overline{T}^{\prime}V_1^{\prime}\right)
  T^{\prime \frac{t-x}{2}-1}\ket{r^{\prime}(t)}}_{\Delta C_r(\frac{t-x}{2}-1,t)},
\end{eqnarray}
where $\Delta C_l(m,t)$ and $\Delta C_r(m,t)$ correspond
to the left and right movers respectively. The contributions $\Delta C_l(m,t)$
and $\Delta C_r(m,t)$ can be evaluated explicitly, see \ref{sec:rhoRhoST}, yielding a following relation
\begin{eqnarray}\label{eq:exactDifferenceRhoRho}
  \Delta C(x,t)=\begin{cases}
    2^{2t+x}\left(
      \binom{-x-3}{\frac{x+t}{2}}
      -\binom{-x-3}{\frac{x+t-2}{2}}
    \right),& x\le -3,\\
    0,& -2 \le  x \le 0,\\
    -2^{2t-x-2}\left(
      \binom{x-1}{\frac{t-x-2}{2}}
      -\binom{x-1}{\frac{t-x-4}{2}}
    \right), & x\ge1.
  \end{cases}
\end{eqnarray}
The correlation function $C(x,t)$ can be obtained recursively from $ \Delta C(x,t) $, and reads
\begin{eqnarray} \label{eq:formalSumSpatioTemporal}
  \eqalign{
    C(x,t)&=C(-t,t)+2^{-2t-1}\sum_{m=0}^{\frac{t-|x|-2}{2}} \Delta C(-t+2m,t)=\\
    &=2^{-t-1}\sum_{m=0}^{\frac{t-|x|-2}{2}} 4^{m}\left(
      2\binom{t-2m-3}{m}-\binom{t-2m-2}{m}
    \right),
  }
\end{eqnarray}
where we took into account that the dynamic structure factor vanishes on the edge of the light cone, $C(-t,t)=0$.
\begin{figure}
  \centering
  \input{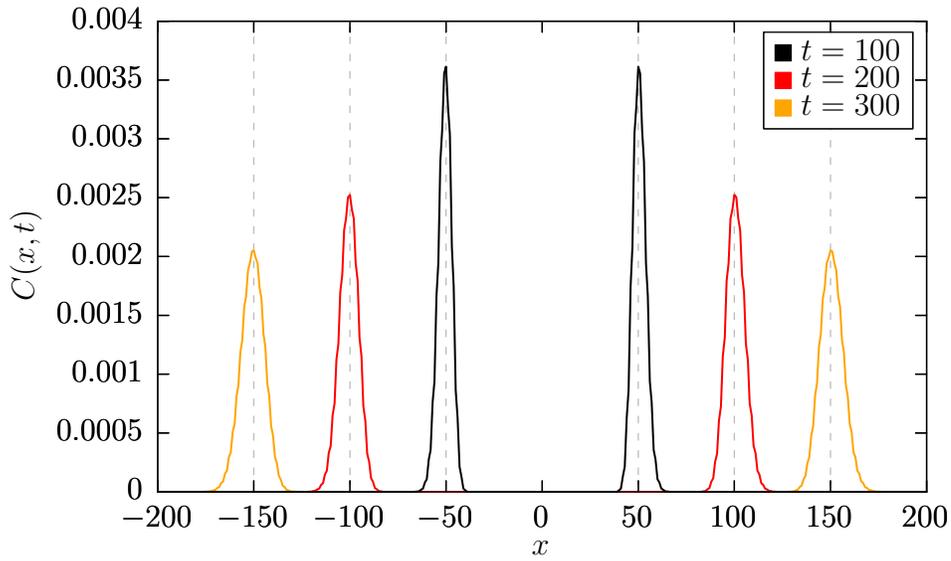}
  \caption{
    \label{fig:rhorhoplot}
    The dynamic structure factor $C(x,t)$ at different $t$. The two peaks move ballistically with the velocity $1/2$,
    while they spread as $\sqrt{t}$.
  }
\end{figure}
Similarly as in the case of the inhomogeneous quench problem, the structure factor can be divided into separate regimes.

\subsection{Homogeneous regime}
This regime occurs in the region $\left|x\right|\le \frac{t}{3}+1$, where the correlation functions for $t\ge 3$ become spatially independent, save for the staggering,
\begin{eqnarray}\label{eq:innerAreaCorrs}
  C(x,t)=2^{-t-1} c_0(t),\\ \nonumber c_0(t)=
  \frac{1}{4}\left(1+\frac{\ii}{\sqrt{7}}\right)
  \left(\frac{-1-\ii \sqrt{7}}{2}\right)^t
  +\frac{1}{4}\left(1-\frac{\ii}{\sqrt{7}}\right)
  \left(\frac{-1+\ii \sqrt{7}}{2}\right)^t.
\end{eqnarray}
This result can be straightforwardly obtained by noting that
the function $s(u)=\sum_{m=0}^{\lfloor\frac{u}{2}\rfloor}4^m\binom{u-2m}{m}$
satisfies the recurrence relation $s(u)=s(u-1)-4 s(u-3)$. Solving the
recurrence relation for the initial conditions $s(0)=s(1)=s(2)=1$, yields the
result~\eqref{eq:innerAreaCorrs}.
Asymptotically, the correlation functions in this regime decrease as $C(x,t)\sim1/\sqrt{2}^t$.

\subsection{Diffusive regime}
\begin{figure}
  \centering
  \input{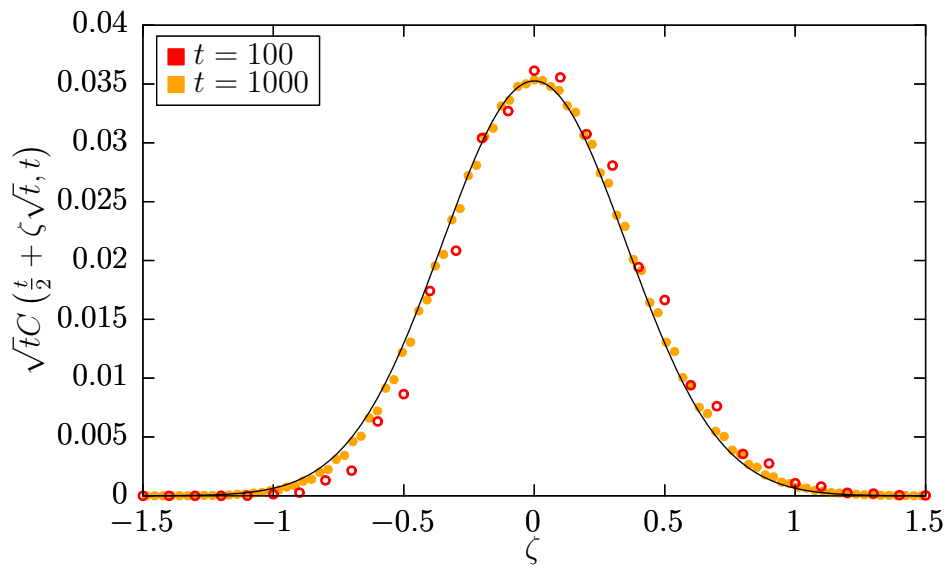}
  \caption{
    \label{fig:rhorhoasymptotics}
    The correlation profile close to the peak. The solid curve denotes the asymptotic shape,
    $\tilde{C}(\zeta)=\frac{1}{16\sqrt{\pi}}\e^{-4\zeta^2}$.
  }
\end{figure}

In the diffusive regime the correlation functions $C(x,t)$ comprise of two asymptotically diffusing peaks, moving apart with a constant velocity $ v=\pm \frac{1}{2} $, see Figure \ref{fig:rhorhoplot}.
\begin{figure}
  \definecolor{inner}{rgb}{1,0.5,0}
\definecolor{inner}{RGB}{254,216,177}
\begin{adjustbox}{scale=0.55,center}
  \begin{tikzpicture}[every node/.style={anchor=center,font=\small,minimum width={25pt},minimum height={25pt},text depth=0ex}]
    \matrix[matrix of nodes](m){
      \phantom{0}& \phantom{0}& \phantom{0}& \phantom{0}& \phantom{0}&
      \phantom{0}& \phantom{0}& \phantom{0}& \phantom{0}& \phantom{0}&
      \phantom{0}& \phantom{0}& \phantom{0}& \phantom{0}&
      \phantom{0}&0&\phantom{0}&0&\phantom{0}&\phantom{0}&\phantom{0}&\phantom{0}&\phantom{0}&\phantom{0}&\phantom{0}&\phantom{0}&\phantom{0}&\phantom{0}&\phantom{0}&\phantom{0}&\phantom{0}&\phantom{0}&\phantom{0}\\
      \phantom{0}& \phantom{0}& \phantom{0}& \phantom{0}& \phantom{0}&
      \phantom{0}& \phantom{0}& \phantom{0}& \phantom{0}& \phantom{0}&
      \phantom{0}& \phantom{0}& \phantom{0}&
      \phantom{0}&0&\phantom{0}&-1&\phantom{0}&0&\phantom{0}&\phantom{0}&\phantom{0}&\phantom{0}&\phantom{0}&\phantom{0}&\phantom{0}&\phantom{0}&\phantom{0}&\phantom{0}&\phantom{0}&\phantom{0}&\phantom{0}&\phantom{0}\\
      \phantom{0}& \phantom{0}& \phantom{0}& \phantom{0}& \phantom{0}&
      \phantom{0}& \phantom{0}& \phantom{0}& \phantom{0}& \phantom{0}&
      \phantom{0}& \phantom{0}&
      \phantom{0}&0&\phantom{0}&1&\phantom{0}&1&\phantom{0}&0&\phantom{0}&\phantom{0}&\phantom{0}&\phantom{0}&\phantom{0}&\phantom{0}&\phantom{0}&\phantom{0}&\phantom{0}&\phantom{0}&\phantom{0}&\phantom{0}&\phantom{0}\\
      \phantom{0}& \phantom{0}& \phantom{0}& \phantom{0}& \phantom{0}&
      \phantom{0}& \phantom{0}& \phantom{0}& \phantom{0}& \phantom{0}&
      \phantom{0}&
      \phantom{0}&0&\phantom{0}&1&\phantom{0}&1&\phantom{0}&1&\phantom{0}&0&\phantom{0}&\phantom{0}&\phantom{0}&\phantom{0}&\phantom{0}&\phantom{0}&\phantom{0}&\phantom{0}&\phantom{0}&\phantom{0}&\phantom{0}&\phantom{0}\\
      \phantom{0}& \phantom{0}& \phantom{0}& \phantom{0}& \phantom{0}&
      \phantom{0}& \phantom{0}& \phantom{0}& \phantom{0}& \phantom{0}&
      \phantom{0}&0&\phantom{0}&1&\phantom{0}&-3&\phantom{0}&-3&\phantom{0}&1&\phantom{0}&0&\phantom{0}&\phantom{0}&\phantom{0}&\phantom{0}&\phantom{0}&\phantom{0}&\phantom{0}&\phantom{0}&\phantom{0}&\phantom{0}&\phantom{0}\\
      \phantom{0}& \phantom{0}& \phantom{0}& \phantom{0}& \phantom{0}&
      \phantom{0}& \phantom{0}& \phantom{0}& \phantom{0}&
      \phantom{0}&0&\phantom{0}&1&\phantom{0}&1&\phantom{0}&1&\phantom{0}&1&\phantom{0}&1&\phantom{0}&0&\phantom{0}&\phantom{0}&\phantom{0}&\phantom{0}&\phantom{0}&\phantom{0}&\phantom{0}&\phantom{0}&\phantom{0}&\phantom{0}\\
      \phantom{0}& \phantom{0}& \phantom{0}& \phantom{0}& \phantom{0}&
      \phantom{0}& \phantom{0}& \phantom{0}&
      \phantom{0}&0&\phantom{0}&1&\phantom{0}&5&\phantom{0}&5&\phantom{0}&5&\phantom{0}&5&\phantom{0}&1&\phantom{0}&0&\phantom{0}&\phantom{0}&\phantom{0}&\phantom{0}&\phantom{0}&\phantom{0}&\phantom{0}&\phantom{0}&\phantom{0}\\
      \phantom{0}& \phantom{0}& \phantom{0}& \phantom{0}& \phantom{0}&
      \phantom{0}& \phantom{0}&
      \phantom{0}&0&\phantom{0}&1&\phantom{0}&9&\phantom{0}&-7&\phantom{0}&-7&\phantom{0}&-7&\phantom{0}&9&\phantom{0}&1&\phantom{0}&0&\phantom{0}&\phantom{0}&\phantom{0}&\phantom{0}&\phantom{0}&\phantom{0}&\phantom{0}&\phantom{0}\\
      \phantom{0}& \phantom{0}& \phantom{0}& \phantom{0}& \phantom{0}&
      \phantom{0}&
      \phantom{0}&0&\phantom{0}&1&\phantom{0}&13&\phantom{0}&-3&\phantom{0}&-3&\phantom{0}&-3&\phantom{0}&-3&\phantom{0}&13&\phantom{0}&1&\phantom{0}&0&\phantom{0}&\phantom{0}&\phantom{0}&\phantom{0}&\phantom{0}&\phantom{0}&\phantom{0}\\
      \phantom{0}& \phantom{0}&\phantom{0}& \phantom{0}& \phantom{0}&
      \phantom{0}&0&\phantom{0}&1&\phantom{0}&17&\phantom{0}&17&\phantom{0}&17&\phantom{0}&17&\phantom{0}&17&\phantom{0}&17&\phantom{0}&17&\phantom{0}&1&\phantom{0}&0&\phantom{0}&\phantom{0}&\phantom{0}&\phantom{0}&\phantom{0}&\phantom{0}\\
      \phantom{0}& \phantom{0}&\phantom{0}& \phantom{0}&
      \phantom{0}&0&\phantom{0}&1&\phantom{0}&21&\phantom{0}&53&\phantom{0}&-11&\phantom{0}&-11&\phantom{0}&-11&\phantom{0}&-11&\phantom{0}&53&\phantom{0}&21&\phantom{0}&1&\phantom{0}&0&\phantom{0}&\phantom{0}&\phantom{0}&\phantom{0}&\phantom{0}\\
      \phantom{0}& \phantom{0}&\phantom{0}&
      \phantom{0}&0&\phantom{0}&1&\phantom{0}&25&\phantom{0}&105&\phantom{0}&-23&\phantom{0}&-23&\phantom{0}&-23&\phantom{0}&-23&\phantom{0}&-23&\phantom{0}&105&\phantom{0}&25&\phantom{0}&1&\phantom{0}&0&\phantom{0}&\phantom{0}&\phantom{0}&\phantom{0}\\
      \phantom{0}& \phantom{0}&
      \phantom{0}&0&\phantom{0}&1&\phantom{0}&29&\phantom{0}&173&\phantom{0}&45&\phantom{0}&45&\phantom{0}&45&\phantom{0}&45&\phantom{0}&45&\phantom{0}&45&\phantom{0}&173&\phantom{0}&29&\phantom{0}&1&\phantom{0}&0&\phantom{0}&\phantom{0}&\phantom{0}\\
      \phantom{0}&
      \phantom{0}&0&\phantom{0}&1&\phantom{0}&33&\phantom{0}&257&\phantom{0}&257&\phantom{0}&1&\phantom{0}&1&\phantom{0}&1&\phantom{0}&1&\phantom{0}&1&\phantom{0}&257&\phantom{0}&257&\phantom{0}&33&\phantom{0}&1&\phantom{0}&0&\phantom{0}&\phantom{0}\\
      \phantom{0}&0&\phantom{0}&1&\phantom{0}&37&\phantom{0}&357&\phantom{0}&677&\phantom{0}&-91&\phantom{0}&-91&\phantom{0}&-91&\phantom{0}&-91&\phantom{0}&-91&\phantom{0}&-91&\phantom{0}&677&\phantom{0}&357&\phantom{0}&37&\phantom{0}&1&\phantom{0}&0&\phantom{0}\\
    };
    \begin{pgfonlayer}{background}
      \node[inner sep=-15pt,fit=(m-1-18)]  (1) {};
      \node[inner sep=-15pt,fit=(m-3-19)]  (3) {};
      \node[inner sep=-15pt,fit=(m-6-20)]  (4) {};
      \node[inner sep=-15pt,fit=(m-9-21)]  (5) {};
      \node[inner sep=-15pt,fit=(m-12-22)] (6) {};
      \node[inner sep=-15pt,fit=(m-15-23)] (7) {};
      \node[inner sep=-15pt,fit=(m-15-11)] (8) {};
      \node[inner sep=-15pt,fit=(m-12-13)] (9) {};
      \node[inner sep=-15pt,fit=(m-9-14)] (10) {};
      \node[inner sep=-15pt,fit=(m-6-15)] (11) {};
      \node[inner sep=-15pt,fit=(m-3-16)] (12) {};
      \node[inner sep=-15pt,fit=(m-1-17)] (14) {};
      \draw[rounded corners,color=orange,very thick,dashed,fill=orange,fill opacity=0.1,inner sep=-15pt]
      (1.north east) |- (3.north east) |- (4.north east) |- (5.north east) |- (6.north east)
      |- (7.north east) |- (8.south west) |- (8.north east) |- (9.north west)
      |- (10.north west) |- (11.north west) |- (12.north west) |- (14.north west)-- cycle;
      \node [rounded corners, very thick, draw=red, rotate fit=45, fit=(m-3-16) (m-15-4)] {};
      \node [rounded corners, very thick, draw=red, rotate fit=45, fit=(m-7-16) (m-15-8)] {};
      \node [rounded corners, very thick, draw=red, rotate fit=45, fit=(m-11-16) (m-15-12)] {};
      \node [rounded corners, very thick, draw=red, rotate fit=45, fit=(m-15-16) (m-15-16)] {};
      \node [rounded corners, very thick, draw=red, rotate fit=45, fit=(m-1-18) (m-15-32)] {};
      \node [rounded corners, very thick, draw=red, rotate fit=45, fit=(m-5-18) (m-15-28)] {};
      \node [rounded corners, very thick, draw=red, rotate fit=45, fit=(m-9-18) (m-15-24)] {};
      \node [rounded corners, very thick, draw=red, rotate fit=45, fit=(m-13-18) (m-15-20)] {};
    \end{pgfonlayer}
  \end{tikzpicture}
\end{adjustbox}
  \caption{\label{fig:rhorhoStruct}
    The explicit values of the correlations, $2^{t+1}C(x,t)$. In the shaded
    inner area, the correlations are homogeneous in $x$ and given by $c_0(t)$. Along the red bordered
    rays, the values are determined by polynomials of order $\frac{1}{2}(t-\left| x \right|) - 1$.
  }
\end{figure}
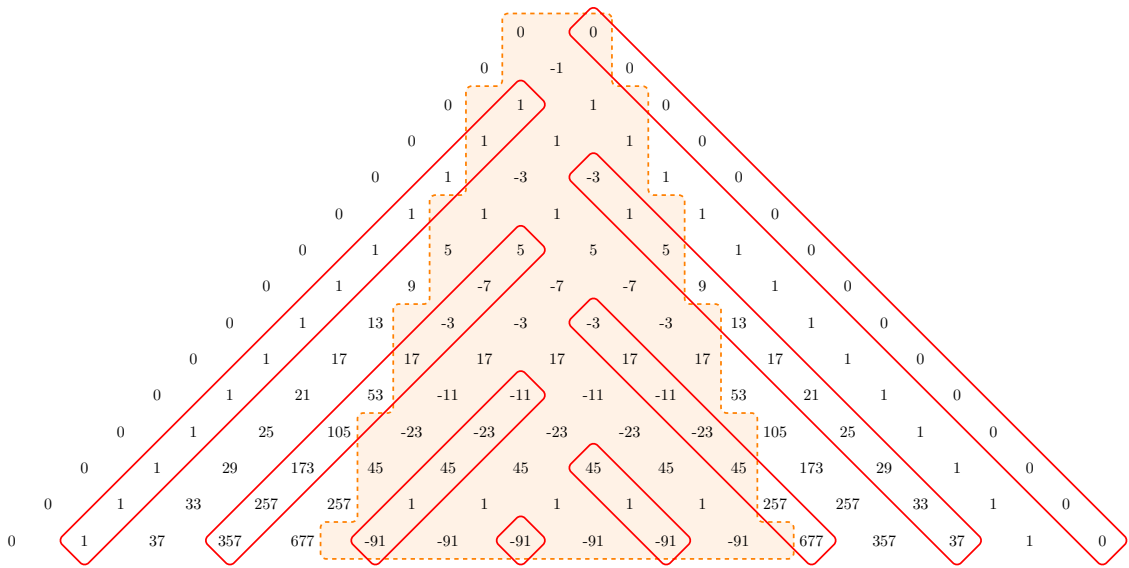
Let us elaborate on an explicit form of the correlation functions in this
regime. Inside of the region $\left|x\right|\ge \frac{t+4}{3}$, the sum
\eqref{eq:formalSumSpatioTemporal} reduces to (see \ref{sec:STcorrsPoly})
\begin{eqnarray}\label{eq:STcorrsPoly}
  C(x,t)=2^{-t-1}
  \smashoperator[l]{\sum_{n=t-\left|x\right|+1}^{\frac{3}{2}\left(t-\left|x\right|\right)}}
  \left(\prod_{\substack{n=t-\left|x\right|+1\\n\neq j}}^{\frac{3}{2}\left(t-\left|x\right|\right)}
  \frac{t-j}{n-j}\right)
  c_0(n).
\end{eqnarray}
Along the rays with the constant distance
from the edge of the light cone, the values of the reduced
correlations, $2^{t+1}C(x,t)$, can be determined by
the polynomial of the order $\frac{1}{2}\left(t-\left|x\right|-2\right)$ in $t$, see Figure \ref{fig:rhorhoStruct}.
In the asymptotic regime, the peaks converge to the normal distribution (see Figure \ref{fig:rhorhoasymptotics}),
\begin{eqnarray}
  C(x,t) \sim
  \frac{1}{16\sqrt{t\pi}}\exp\left(-\frac{4}{t}\left(|x|-\frac{t}{2}\right)^2\right).
\end{eqnarray}
With this we close the discussion of the tMPA construction and its applications.

\section{Conclusions and discussion}\label{sec:conc}
In this article we constructed explicit time evolution of local observables in
terms of the tMPA for a deterministic interacting lattice gas, specifically RCA
54. The auxiliary matrices encode the backward propagation of solitons, where
the auxiliary boundary vectors select the states with a soliton originating
from the center of the chain.

In the second part of the article the tMPA was applied in order to provide
analytical time-dependent solutions of two out-of-equilibrium setups. In the
first one we considered the time-dependent density profile arising from a
piecewise homogeneous initial state,
with the maximum entropy (half-filled) state on the left side of the chain and the empty lattice on the right side of the chain.
In the light-cone around the origin two distinct regions emerge. On the right
side the dynamics reduces to the sea of non-interacting right movers, implying
regime with exponential decay of the density profile with respect to the
spatial coordinate. To the left of the non-interacting sea a thermalizing
region occurs as a consequence of the interactions of solitons, causing an
emergence of the diffusive error function shaped density profile with the
center moving at a constant velocity.

In the second setup we calculated the dynamical structure factor, which is a hallmark of the transport phenomena. 
The structure factor supports two regimes. In the central region a spatially homogeneous state is formed, while the
values outside of the central regime are determined by polynomials along the diagonals. Asymptotically, the correlations behave as
two diffusively spreading Gaussian peaks moving with a constant velocity.
Both of the results analytically demonstrate the coexistence of the ballistic and diffusive transport.

We used the maximum entropy state for our calculations, which is invariant
under time translations, and represents a caricature of a high-temperature
state in physics.
We believe that our explicit computations should be generalizable to a larger
family of invariant clustering states which are described in terms of $4\times 4$
transfer matrices with two free spectral parameters $\xi,\omega$
\cite{prosenMejiaMonasterioCA54,prosenBucaDecayModesCA54},
specifically $p_{\cdots s_{-2},s_{-1},s_0,s_1,s_2,s_3\cdots} = \cdots
T_{s_{-2}s_{-1},s_{0}s_1}(\xi,\omega)T_{s_0 s_1,s_2 s_3}(\omega,\xi)T_{s_2
s_3,s_4 s_5}(\xi,\omega)\cdots$, where $\xi,\omega$ should be real and positive
and
\begin{eqnarray}
T(\xi,\omega) = \begin{bmatrix}
1 & 1 & \xi & 1 \\
\xi \omega & \xi \omega & 1 & \omega \\
\omega & \omega & \xi \omega & \omega \\
\xi & \xi & \xi & \xi \omega
\end{bmatrix}.
\end{eqnarray}
Note that the maximum entropy (separable) state corresponds to $\xi=\omega=1$.

The research presented in this paper opens several interesting questions, both
from the mathematical as well as physical perspective.
\begin{enumerate}
	\item
The first question is what is the type of models for which we can obtain
explicit time dependent representation of local observables in terms of the
tMPA. Namely, the results presented in this paper go beyond what is currently
possible in generic integrable models. At this point we conjecture that such a
solution should be attainable for any purely solitonic, deterministic model
with discrete space-time dynamics. This line of thought is supported by the
tMPA solution of the somewhat simpler model of hard-core interacting charged
particles~\cite{PhysRevLett.119.110603} (see \cite{medenjakPhdThesis}
for an explicit construction of the tMPA). We hope that our approach might offer
some insight into more complicated systems, for example the discrete
space-time versions of the exclusion processes~\cite{Rajewsky1998,VANICAT2018298},
or even space-time discrete  quantum integrable models \cite{VanicatZadnikProsen2017}.
	\item
The second question is how far the explicit tMPA parametrization can push our
knowledge regarding the fluctuations in solitonic models. Ideally, one could
derive a complete large deviation functional for the types of the models
considered in the present paper, and explicitly monitor the validity of the
macroscopic fluctuation theory \cite{bertini2015macroscopic}.
	\item
Furthermore, the origins of, and explicit algebraic structure behind
integrability of the model presented in this paper remain largely unexplored.
This question is naturally linked to the first one.
	\item
Finally, it would be worthwhile studying the tMPA solvable models exhibiting
different types of transport behavior \cite{popkov2015fibonacci,ilievski2018super},
since they could provide the
insight into the microscopic roots of different transport universality classes.
\end{enumerate}

\section*{Acknowledgements}
The work has been supported by Advanced Grant 694544 -- OMNES of European
Research Council (ERC), and by Research program P1-0044 of 
Slovenian Research Agency (ARRS).

\section*{References}
\bibliographystyle{iopart-num}
\bibliography{bobenko}
\appendix

\section{The action of the matrices \texorpdfstring{$T$}{T}, \texorpdfstring{$T^{\prime}$}{T'},
\texorpdfstring{$\overline{T}$}{Tb} and \texorpdfstring{$\overline{T}^{\prime}$}{Tb'}}\label{sec:actionOfT}
In this appendix we explicitly compute the powers of the matrices $T$ and $\overline{T}$ ($T^{\prime}$ and $\overline{T}^{\prime}$)
acting onto the left (right), with the matrices defined as
\begin{eqnarray}
  \eqalign{
    T^{\phantom{\prime}}=(V_0+V_1)(W_0+W_1),\qquad
    &\overline{T}^{\phantom{\prime}}=(W_0+W_1)(V_0+V_1),\\
    T^{\prime} = (W_0^{\prime}+W_1^{\prime})(V_0^{\prime}+V_1^{\prime}),
    &\overline{T}^{\prime} = (V_0^{\prime}+V_1^{\prime})(W_0^{\prime}+W_1^{\prime}),
  }
\end{eqnarray}
but first let us discuss the general structure of the matrices
$M\in \{\alpha V_0+\beta V_1,\alpha W_0+\beta W_1;\,\alpha,\beta\in\mathbb{R}\}$ and
$M^{\prime}\in \{\alpha V^{\prime}_0+\beta V^{\prime}_1,\alpha W^{\prime}_0+\beta W^{\prime}_1;\,\alpha,\beta\in\mathbb{R}\}$.
We start by defining the projectors to the subspace of ``unactivated'' vectors, i.e.\ the subspace
defined by $a=0$, the subspace of ``activated'' vectors with width $0$ ($a=1$ and $w=0$) and to the subspace
of vectors with $a=1$ and $w\le 1$,
\begin{eqnarray}
  \nonumber
  P_0 \ket{c,w,n,a} &= \delta_{a,0}\ket{c,w,n,0},\\
  Q^{\phantom{\prime}}\ket{c,w,n,a} &= \delta_{a,1}\delta_{w,0} \ket{c,0,n,1},\\
  \nonumber
  Q^{\prime}\ket{c,w,n,a} &= \delta_{a,1}\delta_{w,0} \ket{c,0,n,1}
  +\delta_{a,1}\delta_{w,1} \ket{c,1,n,1}.
\end{eqnarray}
The subspace with $a=1$ is invariant to multiplication by matrices $M^{T}$ and $M^{\prime}$,
\begin{eqnarray}
  P_0 M^T P_0 = P_0 M^T,\qquad
  &P_0 M^{\prime} P_0 = P_0 M^{\prime},
\end{eqnarray}
and the value of $w$ inside the $a=1$ subspace cannot increase, which implies the following,
\begin{eqnarray}
  \eqalign{
    Q M^T Q = M^T Q ,\qquad
    &Q M^{\prime} Q = M^{\prime} Q ,\\
    Q^{\prime} M^T Q^{\prime} = M^T Q^{\prime} ,\qquad
    &Q^{\prime} M^{\prime} Q^{\prime} = M^{\prime} Q^{\prime}.
  }
\end{eqnarray}
Additionally, the matrices $M$ ($M^{\prime}$) commute with the raising/lowering
operators defined in the main text equation~\eqref{eq:DefLadderOperators} as long as $w\ge 1$ ($w\ge 2$).
Explicitly,
\begin{eqnarray}
  \eqalign{
    (1-Q)\left[\mathbf{c}^{\pm},M^T\right]=0,
    \qquad
    &
    (1-Q)\left[\mathbf{w}^{\pm},M^T\right]=0,
    \\
    (1-Q^{\prime})\left[\mathbf{c}^{\pm},M^{\prime}\right]=0,
    \qquad
    &
    (1-Q^{\prime})\left[\mathbf{w}^{\pm},M^{\prime}\right]=0.
  }
\end{eqnarray}

We wish to obtain $\bra{v}T^x$, $\bra{v}\overline{T}^{x}$ 
(or $T^{\prime\,x}\ket{v}$, $\overline{T}^{\prime\,x}\ket{v}$)
for an arbitrary vector $\ket{v}\in\mathcal{V}$. Due to the mentioned
properties, it is convenient to first express the $w\ge 1$ (or $w\ge 2$) projections,
\begin{eqnarray}
  \eqalign{
    \bra{v}T^x(1-Q),\qquad
    &\bra{v}\overline{T}^x(1-Q),\\
    (1-Q^{\prime})T^{\prime\,x}\ket{v},
    &(1-Q^{\prime})\overline{T}^{\prime\,x}\ket{v},
  }
\end{eqnarray}
and compute the relevant overlaps using these
vectors (for specific examples see
equations~\eqref{eq:overlapFormalSumL}, \eqref{eq:overlapFormalSumR}, \eqref{eq:dClFormalSum} and \eqref{eq:dCrFormalSum}).

Furthermore, the matrices $V_s^T$, $W_s^T$ and $V_s^{\prime}$, $W_s^{\prime}$
differ only in the boundary terms; explicitly,
\begin{eqnarray}
  (1-Q^{\prime})(V_s^{\prime}-V_s^{T})=0,\qquad
  (1-Q^{\prime})(W_s^{\prime}-W_s^{T})=0,
\end{eqnarray}
therefore we can express the products of right-soliton matrices by projecting the corresponding
left-soliton products to the subspace with $w\ge 2$ and transpose them,
\begin{eqnarray}\label{eq:transposeAndProject}
  \eqalign{
    (1-Q^{\prime})T^{\prime\, x}\ket{v} = \Big(\bra{v}T^x (1-Q)(1-Q^{\prime})\Big)^T,\\
    (1-Q^{\prime})\overline{T}^{\prime\, x}\ket{v} = \Big(\bra{v}\overline{T}^x (1-Q)(1-Q^{\prime})\Big)^T.
  }
\end{eqnarray}
Thus, it suffices to express $\bra{v} T^x(1-Q)$ and $\bra{v} \overline{T}^x(1-Q)$.

\subsection{The powers $T^m$}
The matrices $W_s$, $V_s$ restricted to the subspace with $a=0$ are simple, as are $T$ and $\overline{T}$,
\begin{eqnarray}\fl
  \bra{c,w,n,0} T P_0 &=\bra{c,w,n,0} \overline{T} P_0 = \\ \nonumber
  &=2 \bra{c+1,w-1,0,0} + \bra{c+1,w-1,1,0}+\bra{c+1,w-1,2,0}.
\end{eqnarray}
Since the subspace with $a=1$ is an invariant subspace of the left action of matrices $V_s$, $W_s$,
the following holds
\begin{eqnarray}\fl
  \label{eq:propNoact}
  \bra{c,w,n,0} T^x P_0\!&= \!
  \bra{c,w,n,0} \overline{T}^x P_0 =\\ \nonumber \fl
  &=\!4^{x-1}\!\left(2 \bra{c+x,w-x,0,0} + \bra{c+x,w-x,1,0}+\bra{c+x,w-x,2,0}\right),
\end{eqnarray}
as long as $x\le w$, otherwise the r.h.s.\ is $0$.

Now let us focus on the subspace spanned by $\{\ket{c,w,n,1}; c\ge0, w>0, n\in\{0,1,2\}\}$.
Due to the value of $w$ and $c$
decreasing, it is convenient to express the left action of $T^x$ to the basis
vectors $\bra{c,w,n,1}$ in the following form
\begin{eqnarray}\label{eq:propAct}
  \eqalign{
    \bra{c,w,n,1} T^x (1-Q)
    &= \sum_{m,p} f_{x}^n(m,p) \bra{c-m,w-x-p,0,1}+\\
    &+ \sum_{m,p} g_{x}^n(m,p) \bra{c-m,w-x-p,1,1}+\\
    &+ \sum_{m,p} h_{x}^n(m,p) \bra{c-m,w-x-p,2,1},
  }
\end{eqnarray}
where $f_x^n$, $g_x^n$, $h_x^n$ are some unknown coefficients that have to satisfy the following
recurrence relation
\begin{eqnarray}
  \nonumber
  \eqalign{
    f^n_{x+1}(m,p)&=f^n_x(m,p)+f^n_x(m-1,p-1)+g^n_x(m,p-1)+\\
    &+g^n_x(m-1,p-1)+h^n_x(m,p)+h^n_x(m,p-1),
  }\\
  \label{eq:recurrenceNoBar}
  g^n_{x+1}(m,p)=f^n_x(m,p)+g^n_x(m-1,p-1)+h^n_x(m,p),\\
  \nonumber
  h^n_{x+1}(m,p)=f^n_x(m-1,p)+g^n_x(m-1,p-1)+h^n_x(m-1,p-1).
\end{eqnarray}
A family of solutions is parametrized by $4$ parameters, $\alpha$, $\beta$, $\gamma$ and $\delta$,
\begin{eqnarray}
  \nonumber
  f_x(m,p)=\binom{x-m+p+\alpha}{m+\beta}\binom{x+m-p+\gamma}{p+\delta},\\
  \label{eq:solutionNoBar}
  g_x(m,p)=\binom{x-m+p+\alpha}{m+\beta}\binom{x+m-p-1+\gamma}{p+\delta},\\
  \nonumber
  h_x(m,p)=\binom{x-m+p+\alpha}{m-1+\beta}\binom{x+m-p-1+\gamma}{p+\delta}.
\end{eqnarray}
Taking into account the appropriate initial conditions, it is possible to express
the coefficients $f_x^n$, $g_x^n$, $h_x^n$ in terms of this solution with the following
parametrization,
\begin{eqnarray}
  \nonumber
  n=0:\qquad (\alpha,\beta,\gamma,\delta)=(0,0,0,0),\\
  \label{eq:parametrizationNoBar}
  n=1:\qquad (\alpha,\beta,\gamma,\delta)=(0,0,0,-1),\\
  \nonumber
  n=2:\qquad (\alpha,\beta,\gamma,\delta)=(-1,0,1,0).
\end{eqnarray}
Now we are almost able to express the whole $\bra{v}T^{x}(1-Q)$ for any
vector $\bra{v}$. The last remaining property is
\begin{eqnarray}
  \nonumber
  \bra{c,w,0,0} T (1-P_0)=\bra{c,w-1,0,1},\\
  \label{eq:prodActiv}
  \eqalign{
    \bra{c,w,1,0} T (1-P_0)&=\bra{c,w-1,0,1}+\bra{c-1,w-1,0,1}+\\
    &+\bra{c-1,w-1,1,1}+\bra{c-1,w-1,2,1},
  }\\
  \nonumber
  \bra{c,w,2,0} T (1-P_0)=\bra{c,w-1,0,1}+\bra{c-1,w-1,2,1}.
\end{eqnarray}
Combining the equations~\eqref{eq:propNoact} and~\eqref{eq:prodActiv} with
the expressions~\eqref{eq:solutionNoBar}, and~\eqref{eq:parametrizationNoBar},
we can explicitly obtain the coefficients in the basis expansion of $\bra{v}T^x(1-Q)$
in terms of sums of coefficients~\eqref{eq:solutionNoBar}. For sufficiently
simple $\bra{v}$ they simplify, as for example in the case $\bra{v}=\bra{0,t,0,0}=\bra{l(t)}$,
\begin{eqnarray}
  \label{eq:Tto00}
  \fl\eqalign{
    \bra{l(t)}T^x(1-Q)
    &= 4^{x-1}\left(2\bra{x,t-x,0,0}+\bra{x,t-x,1,0}+\bra{x,t-x,2,0}\right)+\\
    &+\sum_{m=0}^x\smashoperator[r]{\sum_{p=0}^{\min\{m-1,t-x-1\}}}\ 
    \mathcal{A}_x^0(m,p)\bra{x-m,t-x-p,0,1}+\\
    &+\sum_{m=0}^x\smashoperator[r]{\sum_{p=0}^{\min\{m-2,t-x-1\}}}\ 
    \mathcal{A}_x^1(m,p)\bra{x-m,t-x-p,1,1}+\\
    &+\sum_{m=0}^x\smashoperator[r]{\sum_{p=0}^{\min\{m-2,t-x-1\}}}\ 
    \mathcal{A}_x^2(m,p)\bra{x-m,t-x-p,2,1};
  }\\ \label{eq:defA} \fl
  \mathcal{A}_x^0(m,p)=2^{2x+p-m-1}\binom{m-p-1}{p},\qquad
  \mathcal{A}_x^{1,2}(m,p)=2^{2x+p-m-1}\binom{m-p-2}{p}.
\end{eqnarray}

\subsection{The powers $\overline{T}^m$}
Similarly, the left action of $\overline{T}^x$ on $\bra{c,w,n,1}$ can be expressed
in terms of basis vectors via coefficients $\bar{f}_x^n$, $\bar{g}_x^n$, $\bar{h}_x^n$
as
\begin{eqnarray}\label{eq:propbarAct}
  \eqalign{
    \bra{c,w,n,1} T^x (1-Q)
    &= \sum_{m,p} \bar{f}_{x}^n(m,p) \bra{c-m,w-x-p,0,1}+\\
    &= \sum_{m,p} \bar{g}_{x}^n(m,p) \bra{c-m,w-x-p,1,1}+\\
    &= \sum_{m,p} \bar{h}_{x}^n(m,p) \bra{c-m,w-x-p,2,1},
  }
\end{eqnarray}
with the coefficients satisfying a recurrence relation similar to~\eqref{eq:recurrenceNoBar},
\begin{eqnarray}
  \nonumber
  \eqalign{
    \bar{f}^n_{x+1}(m,p)&=\bar{f}^n_x(m,p)+\bar{f}^n_x(m-1,p-1)+\bar{g}^n_x(m-1,p)+\\
    &+\bar{g}^n_x(m-1,p-1)+\bar{h}^n_x(m,p)+\bar{h}^n_x(m-1,p),
  }\\
  \label{eq:recurrenceBar}
  \bar{g}^n_{x+1}(m,p)=\bar{f}^n_x(m,p)+\bar{g}^n_x(m-1,p-1)+\bar{h}^n_x(m,p),\\
  \nonumber
  \bar{h}^n_{x+1}(m,p)=\bar{f}^n_x(m,p-1)+\bar{g}^n_x(m-1,p-1)+\bar{h}^n_x(m-1,p-1).
\end{eqnarray}
Again, a family of solutions is parametrized by $4$ parameters,
\begin{eqnarray}
  \nonumber
  \bar{f}_x(m,p)=\binom{x-m+p+\alpha}{m+\beta}\binom{x+m-p+\gamma}{p+\delta},\\
  \label{eq:solutionBar}
  \bar{g}_x(m,p)=\binom{x-m+p-1+\alpha}{m+\beta}\binom{x+m-p+\gamma}{p+\delta},\\
  \nonumber
  \bar{h}_x(m,p)=\binom{x-m+p-1+\alpha}{m+\beta}\binom{x+m-p+\gamma}{p-1+\delta},
\end{eqnarray}
and the values of parameters corresponding to particular solutions $\bar{f}^n_x$, $\bar{g}^n_x$, $\bar{h}^n_x$ are
\begin{eqnarray}
  \nonumber
  n=0:\qquad (\alpha,\beta,\gamma,\delta)=(0,0,0,0),\\
  \label{eq:parametrizationBar}
  n=1:\qquad (\alpha,\beta,\gamma,\delta)=(0,-1,0,0),\\
  \nonumber
  n=2:\qquad (\alpha,\beta,\gamma,\delta)=(1,0,-1,0).
\end{eqnarray}
The relation equivalent to~\eqref{eq:prodActiv} is
\begin{eqnarray}
  \nonumber\fl
  \bra{c,w,0,0} \overline{T} (1-P_0)=\bra{c,w-1,0,1}+\bra{c+1,w-1,2,1},\\
  \label{eq:prodActivBar}\fl
  \bra{c,w,1,0} \overline{T} (1-P_0)=
  \bra{c,w-1,0,1}+\bra{c,w-1,1,1}+\bra{c+1,w-1,2,1},\\
  \nonumber\fl
  \bra{c,w,2,0} \overline{T} (1-P_0)=\bra{c+1,w-1,2,1}.
\end{eqnarray}
As before, it is possible to explicitly express $\bra{v} \overline{T}^x$
in terms of sums of coefficients $f^n_x$, $g^n_x$, $h^n_x$
for any vector $\bra{v}$. For some special vectors, the expressions
are simple. For example,
\begin{eqnarray}\fl
  \label{eq:Tbartocw}
  \eqalign{
    \bra{c,w,0,0}\overline{T}^x (1-Q)&=4^{x-1}\big(2\bra{c+x,w-x,0,0}+\bra{c+x,w-x,1,0}+\\
    &+\bra{c+x,w-x,2,0}+\bra{c+x,w-x,2,1} \big)+\\
    &+\smashoperator[l]{\sum_{m=1}^{\min\{c+x-1,2x-1\}}}\smashoperator[r]{\sum_{p=0}^{\min\{w-x-1,m-1\}}}
    \bar{\mathcal{A}}_x^0(m,p)\bra{c+x-m,w-x-p,0,1}+\\
    &+\smashoperator[l]{\sum_{m=1}^{\min\{c+x-1,2x-1\}}}\smashoperator[r]{\sum_{p=0}^{\min\{w-x-1,m-1\}}}
    \bar{\mathcal{A}}_x^1(m,p)\bra{c+x-m,w-x-p,1,1}+\\
    &+\smashoperator[l]{\sum_{m=1}^{\min\{c+x-1,2x-1\}}}\smashoperator[r]{\sum_{p=1}^{\min\{w-x-1,m-1\}}}
    \bar{\mathcal{A}}_x^2(m,p)\bra{c+x-m,w-x-p,2,1}
  }
\end{eqnarray}
with the coefficients $\bar{\mathcal{A}}_x^n$ defined as
\begin{eqnarray}
  \label{eq:defAbar}
  \eqalign{
    \bar{\mathcal{A}}_x^0(m,p)=\binom{m-p-1}{p}\sum_{y=m-x}^{2x+p-m-1}\binom{2x+p-m-1}{y},\\
    \bar{\mathcal{A}}_x^1(m,p)=\binom{m-p-1}{p}\sum_{y=m-x}^{2x+p-m-2}\binom{2x+p-m-2}{y},\\
    \bar{\mathcal{A}}_x^2(m,p)=\binom{m-p-1}{p-1}\sum_{y=m-x}^{2x+p-m-2}\binom{2x+p-m-2}{y}.
  }
\end{eqnarray}
Note that we assumed $w\ge x$.

\section{The inhomogeneous quench}\label{sec:appInhQ}
We wish to explicitly obtain the overlaps
\begin{eqnarray}
  \eqalign{
    L(x,t)=\bra{l(t)} T^x V_0 U^{t-x}\ket{r},\\
    R(x,t)=\bra{l^{\prime}}\overline{T}^{\prime\, x} V_0^{\prime} U^{\prime\, t-x}\ket{r^{\prime}(t)},
  }
\end{eqnarray}
with $0\le x \le t$ and
\begin{eqnarray}
  \eqalign{
    T^{\phantom{\prime}}=(V_0+V_1)(W_0+W_1),\qquad &U^{\phantom{\prime}}=W_0 V_0,\\
    \overline{T}^{\prime} = (V_0^{\prime}+V_1^{\prime})(W_0^{\prime}+W_1^{\prime}),
    &U^{\prime}=W_0^{\prime}V_0^{\prime}.
  }
\end{eqnarray}
Let us start with the overlap that corresponds to the left moving solitons.

\subsection{Expressing the overlap \texorpdfstring{$L(x,t)$}{L(x,t)}}
The matrices $V_s$, $W_s$ act trivially on the vectors $\bra{0,0,n,1}$, 
\begin{eqnarray}
  \eqalign{
    \bra{0,0,n,1}V_s=\bra{0,0,n,1}W_s=\bra{0,0,s\cdot \max\{n+1,2\},1},\\
    \bra{0,0,n,1}T=2\bra{0,0,0,1}+\bra{0,0,1,1}+\bra{0,0,2,1},
  }
\end{eqnarray}
therefore we can treat these vectors separately. Since they are the only
vectors with the nonzero overlap with $\ket{r}$, computing $L(x,t)$ is equivalent
to summing up the contributions of $\bra{0,0,n,1}$ vectors that are created
at different steps. Explicitly,
\begin{eqnarray}\fl\label{eq:overlapFormalSumL}
  \eqalign{
    L(x,t)&=\underbrace{(1-\delta_{x,0})\bra{l(t)}T\ket{r}}_{\equiv L_1(x,t)}+
    \underbrace{\sum_{y=1}^{x} 4^{x-y} \bra{l(t)}T^{y-1}(1-Q) T \ket{r}}_{\equiv L_2(x,t)}+\\
    &+ \underbrace{\bra{l(t)}T^x(1-Q)V_0 \ket{r} + \sum_{z=1}^{t-x-1}
    \bra{l(t)}T^x V_0 U^{z-1} (1-Q) U \ket{r}}_{\equiv L_3(x,t)}.
  }
\end{eqnarray}
Since $L(0,t)=0$, let us from now on assume $x>0$ to simplify the notation.
The first contribution is easy; if $t\neq 0$, the only nonzero overlap occurs
for $x=t=1$,
\begin{eqnarray}
  L_1(x,t)=\delta_{x,1}\delta_{t,1}.
\end{eqnarray}
The second contribution $L_2(x,t)$ is obtained from~\eqref{eq:Tto00} as
\begin{eqnarray}
  \fl
  \eqalign{
    &L_2(x,t) = 4\delta_{x,2}\delta_{t,2}
    +\smashoperator{\sum_{y=\lceil\frac{t+2}{2}\rceil}^{x-1}}
    2\cdot 4^{x-y-1}\mathcal{A}_{y}^1(y-1,t-y-1)+\\
    &+\smashoperator{\sum_{y=\lceil\frac{t+1}{2}\rceil}^{x-1}}
    4^{x-y-1}\left(
      2\mathcal{A}_{y}^0(y-1,t-y-1)
      +6\mathcal{A}_{y}^1(y,t-y-1)
      +4\mathcal{A}_{y}^1(y-1,t-y-2)
    \right)+\\
    &+\smashoperator{\sum_{y=\lceil\frac{t}{2}\rceil}^{x-1}}
    4^{x-y-1}\left(
      2\mathcal{A}_y^0(y,t-y-1)
      +\mathcal{A}_{y}^0(y-1,t-y-2)
      +2\mathcal{A}_{y}^1(y,t-y-2)
    \right).
  }
\end{eqnarray}
Taking into account the form of the coefficients from~\eqref{eq:Tto00} the first
two contributions combine into
\begin{eqnarray}\fl\nonumber
  L_1(x,t)+L_2(x,t)=2^{2x-t-1}\bigg(
    \smashoperator[r]{\sum_{z=t-x}^{\lfloor\frac{t-1}{2}\rfloor}}
    4^z \binom{t-1-2z}{z}+
    2 \smashoperator{\sum_{z=t-x}^{\lfloor\frac{t-2}{2}\rfloor}}
    4^z \binom{t-2-2z}{z}+\\
    \fl\label{eq:L1L2ns}
    \phantom{L_1(x,t)+L_2(x,t)\,}
    + 3 \smashoperator{\sum_{z=t-x}^{\lfloor\frac{t-3}{2}\rfloor}}
    4^z \binom{t-3-2z}{z}+
    2 \smashoperator{\sum_{z=t-x}^{\lfloor\frac{t-4}{2}\rfloor}}
    4^z \binom{t-4-2z}{z}
  \bigg)=\\
  \fl\nonumber
  =
  2^{2x}\left(
    \frac{1}{4}u(t-1,t-x) + \frac{1}{4} u (t-2,t-x) + \frac{3}{16} u(t-3,t-x) +\frac{1}{16} u (t-4,t-x)
  \right);
  \\ \nonumber \fl
  u(m,n) \equiv \sum_{y=n}^{\lfloor\frac{m}{2}\rfloor} 2^{-(m-2y)}\binom{m-2y}{y}.
\end{eqnarray}
The function $u(m,n)$ satisfies the following recurrence relation,
\begin{eqnarray}\fl
  u(m,n) = \frac{1}{2} u(m-1,n) + \frac{1}{2} u(m-3,n) + \theta_{m-2x-1} 2^{2n-m} \binom{m-2n}{n-1},
\end{eqnarray}
which implies
\begin{eqnarray}\fl
  2 u(m,n) + u(m-1,n) + u(m-2,n) =
  \begin{cases}
    2; & n=0,\,m\ge0,\\
    0; & n=0,\,m<0,\\
    \sum_{y=0}^{m-2n-1} 2^{-y}\binom{y}{n-1}; & n>0.
  \end{cases}
\end{eqnarray}
Therefore, the expression~\eqref{eq:L1L2ns}  simplifies into
\begin{eqnarray}
  \nonumber\fl
  L_1(x,t)+L_2(x,t)&=2^{2x} \delta_{t,x}(1-\delta_{t,1})\frac{3}{8}+
  \delta_{t,1}\delta_{x,1}+8\delta_{x,3}\delta_{t,4}
  +2^{2x}
  \smashoperator{\sum_{y=0}^{2x-t-4}} 2^{-y}\binom{y}{t-x-1}
  +\\ \fl \label{eq:L1L2}
  &+\theta_{2x-t-3} 2^{t-1}
  \left(2\binom{2x-t-3}{t-x-1}+\binom{2x-t-3}{t-x-2}\right),
\end{eqnarray}
where $\theta_x$ is a discrete Heaviside function,
\begin{eqnarray}
  \theta_x=\begin{cases}1;& x\ge 0,\\ 0; & x<0.\end{cases}
\end{eqnarray}

The other part is obtained by observing
\begin{eqnarray}
  \bra{c,w,0,1}U^z(1-Q)U\ket{r}=\delta_{c,0}\delta_{w,z+1},
\end{eqnarray}
which implies
\begin{eqnarray}
  \eqalign{
    L_3(x,t)&=\delta_{x,1}
    +\smashoperator{\sum_{z=\max\{1,t-2x+1\}}^{t-x}}\mathcal{A}_x^0(x,t-x-z)
    +\smashoperator{\sum_{z=\max\{1,t-2x+2\}}^{t-x}}\mathcal{A}_x^1(x,t-x-z)+\\
    &+\smashoperator{\sum_{z=\max\{1,t-2x+3\}}^{t-x}}\mathcal{A}_x^1(x-1,t-x-z).
  }
\end{eqnarray}
Inserting the explicit forms of the coefficients $\mathcal{A}_x^n$ and simplifying
the expression we obtain
\begin{eqnarray}\label{eq:L3}
  \eqalign{
    L_3(x,t)&=\delta_{x,1}\delta_{t,1}+\theta_{2x-t-2} 2^{t-1}\binom{2x-t-2}{t-x-1}+\\
    &+2^x\bigg(
      \underbrace{\sum_{y=0}^{x-1} 2^y \binom{x-1-y}{y}}_{\frac{2^x}{3}\left({1-\left(\frac{1}{2}\right)^x}\right)}
      -\sum_{y=t-x}^{x-1}2^y \binom{x-1-y}{y}
    \bigg).
  }
\end{eqnarray}

Finally, the whole contribution of the left MPA is 
\begin{eqnarray}\fl \label{eq:wholeL}
  \eqalign{
    L(x,t)&=
    3\cdot 2^{2x-3} \delta_{t,x}(1-\delta_{t,1})
    +2\delta_{x,1}\delta_{t,1}
    +16\delta_{x,3}\delta_{t,4}
    +2^{2x} \sum_{y=0}^{2x-t-4} 2^{-y}\binom{y}{t-x-1}+\\
    &+2^{2x}\frac{{1-\left(\frac{1}{2}\right)^x}}{3}
    -2^x \sum_{y=t-x}^{x-1}2^y \binom{x-1-y}{y}+\\
    &+\theta_{2x-t-3} 2^{t-1}
    \left(3\binom{2x-t-3}{t-x-1}+2\binom{2x-t-3}{t-x-2}\right).
  }
\end{eqnarray}

\subsection{Overlap \texorpdfstring{$R(x,t)$}{R(x,t)}}
We start by observing
\begin{eqnarray}
  V_0^{\prime} U^{\prime\, t-x}\ket{r^{\prime}(t)}
  =V_0^{\prime}U^{\prime\, t-x}\ket{0,t+1,0,0}=\ket{t-x,x+1,0,0},
\end{eqnarray}
therefore the contribution of right moving solitons to the density profile is
\begin{eqnarray}\label{eq:overlapFormalSumR}
  R(x,t)=\bra{l^{\prime}} \overline{T}^{\prime\, x}V_0^{\prime}U^{\prime\, t-x}\ket{r^{\prime}(t)}
  =\bra{l^{\prime}}\overline{T}^{\prime\, x}\ket{t-x,x+1,0,0}.
\end{eqnarray}
We use the same approach as before; as soon as vectors $\ket{0,0,n,1}$ are created, we compute their
overlap with the left boundary vector, while we keep propagating the other vectors,
\begin{eqnarray}\fl
  \eqalign{
    R(x,t)&=\bra{l^{\prime}}\overline{T}^{\prime} (1-Q) \overline{T}^{\prime\, x-1}\ket{t-x,x+1,0,0}+\\
    &+\sum_{y=1}^{x-2}4^{x-y-1}\bra{r} \overline{T}^{\prime}(1-Q)\overline{T}^{\prime\, y}\ket{t-x,x+1,0,0}+
    \bra{r}\overline{T}^{\prime}\ket{t-x,x+1,0,0},
  }
\end{eqnarray}
where we used the fact that $\bra{r}=(\ket{r})^T$ is the $w=0$ part of $\bra{l^{\prime}}$, i.e.\
\begin{eqnarray}
  \bra{r}=\bra{l^{\prime}}-\left(\bra{0,1,0,1}+\bra{0,1,2,1}\right)=\bra{l^{\prime}}Q.
\end{eqnarray}
The expression for $(1-Q^{\prime})\overline{T}^{\prime\, y}\ket{t-x,x+1,0,0}$ is straightforwardly
obtained from equation~\eqref{eq:Tbartocw} by transposing it and removing the vectors with $w=1$,
therefore the right overlap reads
\begin{eqnarray}\fl
  \eqalign{
    R(x,t)&=2\delta_{t,3}\delta_{x,3}+
    (1-\delta_{t,1})\mathcal{A}_{x-1}^0(t-1,0) \\
    &+\smashoperator{\sum_{y=\max\{1,\lceil\frac{2x-t}{2}\rceil\}}^{x-1}}
    4^{x-y-1}\overline{\mathcal{A}}_{y}^0(t-x+y,x-y-1)+\\
    &+\smashoperator{\sum_{y=\max\{1,\lceil\frac{2x-t+1}{2}\rceil\}}^{x-1}}
    2\cdot 4^{x-y-1}\overline{\mathcal{A}}_{y-1}^0(t-x+y-1,x-y-1)+\\
    &+\smashoperator{\sum_{y=\max\{1,\lceil\frac{2x-t+2}{2}\rceil\}}^{x-1}}
    4^{x-y-1}\bigg(
      4\overline{\mathcal{A}}_{y-1}^0(t-x+y-1,x-y)+
      6\overline{\mathcal{A}}_{y-1}^1(t-x+y-2,x-y-1)+\\
      &+4\overline{\mathcal{A}}_{y-1}^2(t-x+y-1,x-y)+
      2\overline{\mathcal{A}}_{y-1}^2(t-x+y-2,x-y-1)
    \bigg).
  }
\end{eqnarray}
Due to the coefficients inside the sum vanishing for almost all values of $y$,
the overlap can be equivalently expressed as
\begin{eqnarray}\fl
  \eqalign{
    R&(x,t)=2\delta_{t,3}\delta_{x,3} +\smashoperator{\sum_{y=\lceil\frac{2x-t-1}{2}\rceil}^{\infty}}
    2^{2x-2y-3}\bar{\mathcal{A}}_y^0(t-x+y,x-y-2)+\\
    &+\smashoperator{\sum_{y=\lceil\frac{2x-t}{2}\rceil}^\infty}
    2^{2x-2y-3} \bigg(
      4\bar{\mathcal{A}}_{y}^0(t-x+y,x-y-1)+
      3\bar{\mathcal{A}}_{y}^1(t-x+y-1,x-y-2)+\\
      &+2\bar{\mathcal{A}}_{y}^2(t-x+y,x-y-1)+
      \bar{\mathcal{A}}_{y}^2(t-x+y-1,x-y-2)
    \bigg).
  }
\end{eqnarray}
Inserting the explicit values of $\bar{\mathcal{A}}_y^n$ and simplifying the whole expression
yields
\begin{eqnarray}
  \fl\nonumber
  R(x,t)=2\delta_{x,2}\delta_{t,2}+2\delta_{x,3}\delta_{t,3}
  + 2 \left( s(t-2)+ s (t-3) + 2 s(t-4)\right) \sum_{z=t-x}^{2x-t-3}\binom{2x-t-3}{z}+
  \\\fl\phantom{R(x,t)\,}
  + 2 (1-\delta_{t,2})\left( s(t-2)+ 3 s (t-4) + 4 s(t-6)\right) \sum_{z=t-x-1}^{2x-t-3}\binom{2x-t-3}{z};
  \\ \fl \nonumber
  s(m)=\sum_{z=0}^{\lfloor \frac{m}{2}\rfloor} 4^{z} \binom{m-2z}{z}.
\end{eqnarray}
The function $s(m)$ satisfies the following recurrence relation,
\begin{eqnarray}
  s(m+3)=s(m+2)+4 s(m),
\end{eqnarray}
which together with the initial condition $s(0)=s(1)=s(2)=1$ implies~\footnote{Note that
  this holds only for positive $m-4$. If $m\le 3$, we have to explicitly express
the relevant $s(m)$.}
\begin{eqnarray}\fl
  s(m)+3 s(m-2) + 4 s(m-4) = s(m)+2(m-1)+2s(m-2)=2^m,
\end{eqnarray}
and the whole contribution from the right moving solitons is
\begin{eqnarray}\fl\label{eq:wholeR}
  R(x,t)=2\delta_{x,2}\delta_{t,2} + \theta_{2x-t-3}2^{t-1}\binom{2x-t-3}{t-x-1}
  +2^t\sum_{z=t-x}^{2x-t-3}\binom{2x-t-3}{z}
\end{eqnarray}

\section{The free regime of the inhomogeneous quench}
\label{free_regime}
\begin{figure}
  \centering
  \includegraphics[width=\textwidth]{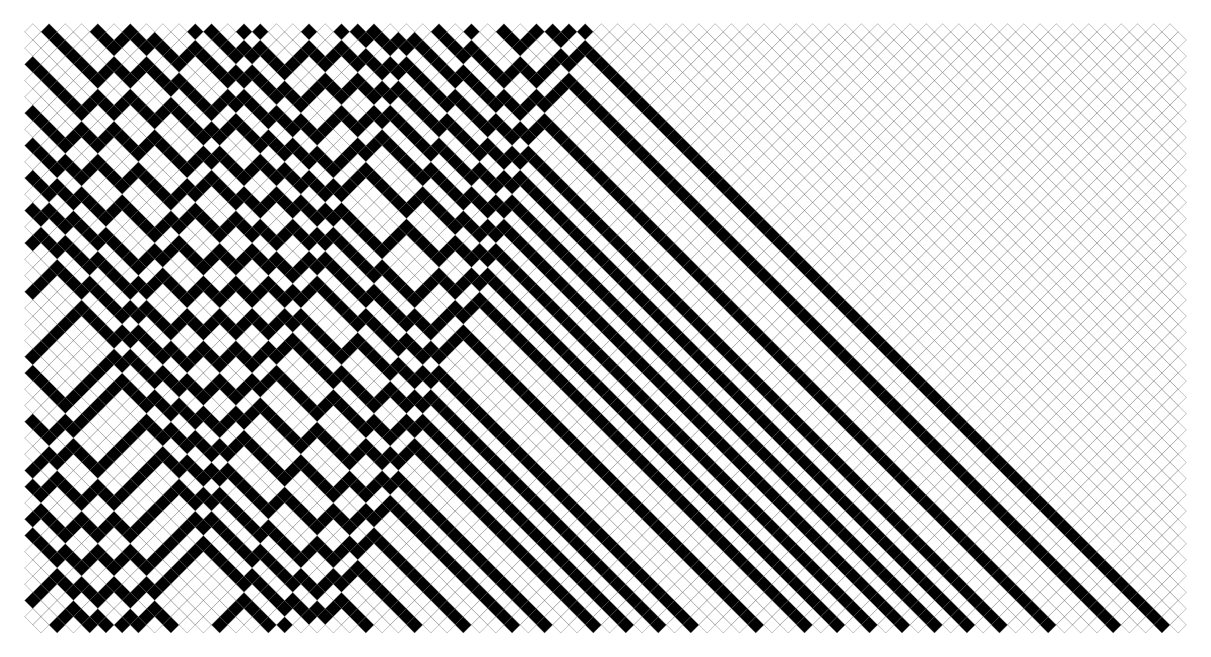}
  \caption{\label{fig:inhQReal}One realization of the inhomogeneous quench up to $t=74$.
    Between the empty space on the right and the area filled with left and right moving solitons, there
    is a section, where all the particles move to the right and do not scatter, corresponding to the free regime.
  }
\end{figure}
In this section we present an alternative derivation of the expression for
the density profile~\eqref{eq:inhQprofileIntermediate},
which also provides some physical insights into the result.
Before the quench, there are no solitons in the right half-infinite chain. When we join
the two half-chains, the right moving solitons from the left that reach the boundary
continue moving to the right unperturbed with the velocity $1$, since there are no
left moving solitons to slow them down. Therefore an intermediate area with only
right moving solitons is established between the vacuum and the part that contains both
types of solitons. This can be seen on an example in Figure \ref{fig:inhQReal}.
Due to the maximal velocity of the solitons being $v_{\text{max}}=1$, this area is limited
to the right by $x=t$.
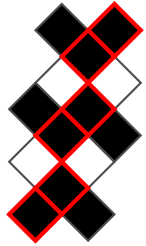
\begin{figure}
  \centering
  \definecolor{full}{rgb}{0,0,0}
\definecolor{old}{rgb}{1,1,1}
\definecolor{border}{rgb}{0.3,0.3,0.3}
\def\a{0.35}
\def\b{8}
\def\off{1.75}
\def\num{3}
\renewcommand\emptyrectangle[2]{
  \draw[thick,border] ({\a*(#1)},{\a*(#2-1)})  -- ({\a*(#1+1)},{\a*(#2)})  -- ({\a*(#1)},{\a*(#2+1)})  
  -- ({\a*(#1-1)},{\a*(#2)})  -- cycle;
}
\renewcommand\fullrectangle[2]{
  \draw[thick,border,fill=full] ({\a*(#1)},{\a*(#2-1)})  -- ({\a*(#1+1)},{\a*(#2)})  -- ({\a*(#1)},{\a*(#2+1)})  
  -- ({\a*(#1-1)},{\a*(#2)})  -- cycle;
}
\renewcommand\halfrectangle[2]{
  \draw[ultra thick,red,fill=full] ({\a*(#1)},{\a*(#2-1)})  -- ({\a*(#1+1)},{\a*(#2)})  -- ({\a*(#1)},{\a*(#2+1)})  
  -- ({\a*(#1-1)},{\a*(#2)})  -- cycle;
}
\begin{tikzpicture}
  \fullrectangle{0}{0};
  \emptyrectangle{0}{-2};
  \emptyrectangle{2}{-2};
  \fullrectangle{-1}{-3};
  \fullrectangle{2}{-4};
  \emptyrectangle{-1}{-5};
  \emptyrectangle{1}{-5};
  \fullrectangle{1}{-7};

  \halfrectangle{2}{0};
  \halfrectangle{1}{-1};
  \halfrectangle{1}{-3};
  \halfrectangle{0}{-4};
  \halfrectangle{0}{-6};
  \halfrectangle{-1}{-7};
\end{tikzpicture}
  \caption{\label{fig:MinSpeed}A path of a soliton (red bordered
    squares) that moves to the left with the effective velocity $1/3$.
    Due to propagation rules, the solitons can not scatter more frequently,
    therefore this is the slowest possible effective speed.
  }
\end{figure}
The left border is determined by the right most possible
position of the left moving solitons, which is $x=-t/3$ due to the effective
soliton speed being bounded from bellow by $1/3$ (see Figure \ref{fig:MinSpeed}).
The ballistic part of the profile is therefore described by the $-t/3+1\le x \le t$
part of the profile in~\eqref{eq:inhQprofileIntermediate}, which can also be derived
by assuming that solitons enter this area randomly with uniform probability.

Let us look at the intermediate area of the chain at some fixed time $t$ and let
us join two consecutive sites together,
so that sites $t-(2k-1)$ and $t-2k$ constitute a \emph{supersite} labeled by $n=k$,
\begin{eqnarray}\label{eq:SuperSites}
  \eqalign{
    \begin{tikzpicture}
      \def\x{0.8/sqrt(2)}
      \def\a{1.5*sqrt(2)}
      \def\off{0.2}
      \nrectangle{0}{0}{$t$};
      \nrectangle{-0.5}{-0.5}{$t-1$};
      \nrectangle{-1}{0}{$t-2$};
      \nrectangle{-1.5}{-0.5}{$t-3$};
      \nrectangle{-2}{0}{$t-4$};
      \nrectangle{-2.5}{-0.5}{$t-5$};
      \nrectangle{-3}{0}{$t-6$};
      \nrectangle{-3.5}{-0.5}{$t-7$};
      \nrectangle{-4}{0}{$t-8$};
      \brectangle{1};
      \brectangle{2};
      \brectangle{3};
      \brectangle{4};
      \node[text=gray] at ({-4.75*\a},{-0.25*\a}) {\LARGE $\mathbf{\cdots}$};
      \node[text=gray] at ({-0.75*\a},{-0.5*\a-0.5*\x-2*\off}) {{$\mathbf{n=1}$}};
      \node[text=gray] at ({-1.75*\a},{-0.5*\a-0.5*\x-2*\off}) {{$\mathbf{n=2}$}};
      \node[text=gray] at ({-2.75*\a},{-0.5*\a-0.5*\x-2*\off}) {{$\mathbf{n=3}$}};
      \node[text=gray] at ({-3.75*\a},{-0.5*\a-0.5*\x-2*\off}) {{$\mathbf{n=4}$}};
    \end{tikzpicture}
  }
\end{eqnarray}
All the sites with $x\ge t$ are empty, while a site with $x\le t-1$ is occupied
if there is a right moving soliton going through it, in which case the whole
supersite has to be occupied and the neighbouring supersites have to be empty.
Since the solitons enter this area randomly, the site $n=1$ is occupied
with probability $1/2$. If site $n-1$ is full, site $n$ has to be empty
and if $n-1$ is empty, site $n$ is occupied with probability $1/2$. This can
be expressed in a matrix form as
\begin{eqnarray}
  \vec{x}_n=\begin{bmatrix} \frac{1}{2} & 1 \\ \frac{1}{2} & 0 \\\end{bmatrix} \vec{x}_{n-1}
  =\begin{bmatrix}\frac{1}{2} & 1\\ \frac{1}{2} & 0\\ \end{bmatrix}^{n-1}\vec{x}_1,\qquad
  \vec{x}_n=\begin{bmatrix}1-p_n\\p_n\end{bmatrix},
\end{eqnarray}
where $p_n$ is the probability of the site $n$ being full. Taking into account $p_1=\frac{1}{2}$,
we obtain
\begin{eqnarray}
  p_n=\frac{1}{3}\left(1-\left(-\frac{1}{2}\right)^n\right),
\end{eqnarray}
which matches the ballistic part of the profile~\eqref{eq:inhQprofileIntermediate}.

\section{The dynamic structure factor}\label{sec:rhoRhoST}
As already discussed in \ref{sec:actionOfT} and \ref{sec:appInhQ}, the matrices
act trivially on the vectors $\bra{0,0,n,1}$, which implies that a general overlap,
$\bra{l(t)}M_1 M_2\cdots M_{2t+1}\ket{r}+
\bra{l^{\prime}}M_1^{\prime} M_2^{\prime}\cdots M_{2t+1}^{\prime}\ket{r^{\prime}(t)}$,
can be straightforwardly determined using the projections
$\bra{l(t)}M_1 M_2 \cdots M_j(1-Q)$, with $j=0,1,\ldots 2t+1$.
Explicitly, the overlaps~\eqref{eq:exactDifferencesLR} can be expressed as
\begin{eqnarray}\label{eq:dClFormalSum} \fl
  \eqalign{
    \Delta C_l(x,t)&=4^{t-x-1}\bra{l(t)}T^x (1-Q) D\ket{r}+
    \smashoperator{\sum_{y=0}^{t-x-2}}
    4^{t-x-y-2}
    \bra{l(t)}T^xD\overline{T}^y (1-Q)
    \overline{T}\ket{r},
  }
\end{eqnarray}
and
\begin{eqnarray}
  \label{eq:dCrFormalSum} \fl\nonumber
  \Delta C_r(x,t)&=4^{t-x-1}\bra{r}
  D^{\prime}
  (1-Q) T^{\prime\, x} \ket{r^{\prime}(t)}+ 
  \smashoperator{\sum_{y=0}^{t-x-2}}4^{t-x-y-2}\bra{r}\overline{T}^{\prime}(1-Q)\overline{T}^{\prime\,y}
  D^{\prime}
  T^{\prime\, x}\ket{r^{\prime}(t)}\\ \fl
  &+\left(\bra{l^{\prime}}-\bra{r}\right)
  \overline{T}^{\prime\,t-x-1}
  D^{\prime}
  T^{\prime\, x}\ket{r^{\prime}(t)},
\end{eqnarray}
where we introduced $D$, $D^{\prime}$ to denote the difference of the matrices,
\begin{eqnarray}
  D=T V_1-V_1 \overline{T},\qquad
  D^{\prime}=V_1^{\prime} \overline{T}^{\prime}-T^{\prime} V_1^{\prime}.
\end{eqnarray}
Therefore to obtain $\Delta C_{l}(x,t)$ it suffices to express the projections
$\bra{l(t)}T^x (1-Q)$, $\bra{l(t)}T^x D (1-Q)$ and $\bra{l(t)}T^{x}D \overline{T}^y(1-Q)$
and then compute the relevant overlaps with the right vector $\ket{r}$
as shown in~\eqref{eq:dClFormalSum}. The right moving soliton counterpart is
very similar; since the matrices $W_s^{\prime}$, $V_s^{\prime}$ are the same
as $W_s^{T}$ and $V_s^T$ in the $w\ge 2$ subspace,
we can just take the corresponding left moving soliton vectors, transpose them,
remove the terms with $w \le 1$ (similarly as in~\eqref{eq:transposeAndProject})
and compute the overlaps from~\eqref{eq:dCrFormalSum}.

The procedure is straightforward but lengthy, therefore we split it into multiple
parts. In \ref{subs:actProjections} we use the relations from~\ref{sec:actionOfT}
to explicitly write the vectors $\bra{l(t)}T^x(1-Q)$, $\bra{l(t)}T^x D (1-Q)$
and $\bra{l(t)}T^x(1-Q)\overline{T}^y$ in terms of basis vectors $\bra{c,w,n,a}$
by introducing the coefficients $\mathcal{A}_x^n$, $\mathcal{B}_x^n$,
$\mathcal{C}_{x,y}^n$ and $\mathcal{D}_{x,y}^n$. In~\ref{subs:differentContributions}
we proceed to express the overlaps $\Delta C_{l,r}(x,t)$. We split the overlaps into
multiple parts corresponding to different coefficients and we simplify the contributions.
They are expressed in terms of single binomial coefficients, their single sums and triple sums.
The contributions consisting of triple sums are simplified in~\ref{subs:deltaDdeltaDprime},
where also the whole overlaps $\Delta C_{l,r}(x,t)$ are expressed.
Additionally, another subsection is included ad the end, \ref{sec:STcorrsPoly}, where we show
the equivalence of expressions~\eqref{eq:formalSumSpatioTemporal}
and~\eqref{eq:STcorrsPoly} from the main text.

\subsection{The explicit form of different contributions to the overlaps}\label{subs:actProjections}
We start by expressing the vectors $\bra{l(t)}T^x(1-Q)$,
$\bra{l(t)}T^x D (1-Q)$
and $\bra{l(t)}T^xD\overline{T}^y(1-Q)$.
The first one can be expressed in terms of the basis vectors $\bra{c,w,n,a}$
with the coefficients $\mathcal{A}_x^n(m,p)$,
as introduced in~\eqref{eq:Tto00} and~\eqref{eq:defA}.
Acting on it with $T V_1 - V_1 \overline{T}$
we straightforwardly obtain
\begin{eqnarray}\label{eq:actWithDiffOnce}\fl
  \eqalign{
    \bra{l(t)}T^x &D(1-Q)=\\
    &\left.
      \eqalign{
        &= 4^x \bigg(-2\bra{x+1,t-(x+1);0;0}
          +\bra{x+1,t-(x+1);0;1} - \\
          &\phantom{= 4^x \bigg(-\ }+\bra{x+1,t-(x+1);0;2}
          +\bra{x+1,t-(x+1);1;2}\bigg)
      }\right\}\bra{s(x,t)}\\\fl
      &\left.
        \eqalign{
          &- \sum_{p}\sum_{m}\mathcal{B}^0_x(m,p)\bra{x-m,t-(x+1)-p;1;0}\\
          &+ \sum_{p}\sum_{m}\mathcal{B}^1_x(m,p)\bra{x-m,t-(x+1)-p;1;1}\\
          &+ \sum_{p}\sum_{m}\mathcal{B}^2_x(m,p)\bra{x-m,t-(x+1)-p;1;2}
      }\right\} \bra{c(x,t)},
    }
  \end{eqnarray}
  with the following coefficients
  \begin{eqnarray}
    \eqalign{
      \mathcal{B}_x^0(m,p)=2^{2x+p-m}\binom{m-p}{p},\quad
      \mathcal{B}_x^1(m,p)=2^{2x+p-m}\binom{m-p-1}{p},\\
      \mathcal{B}_x^2(m,p)=2^{2x+p-m-1}\binom{m-p}{p-1}.
    }
  \end{eqnarray}
  At this point it is convenient to split $\bra{l(t)}T^x D\overline{T}^y(1-Q)$
  into two parts; the first part corresponds to acting with $\overline{T}^y$ from right to the first two lines
  from~\eqref{eq:actWithDiffOnce},
  \begin{eqnarray}
    \eqalign{
      \bra{s(x,t)}\overline{T}^{y}(1-Q)&=
      \sum_{m,p}\mathcal{C}_{x,y}^0(m,p)\bra{x-m,t-(x+y+1)-p;1;0}\\
      &+\sum_{m,p}\mathcal{C}_{x,y}^1(m,p)\bra{x-m,t-(x+y+1)-p;1;1}\\
      &+\sum_{m,p}\mathcal{C}_{x,y}^2(m,p)\bra{x-m,t-(x+y+1)-p;1;2},
    }
  \end{eqnarray}
  where the coefficients $\mathcal{C}_{x,y}^n(m,p)$ are expressed in terms of $\bar{f}_y^n$, $\bar{g}_y^n$
  and $\bar{h}_y^n$ as introduced in~\eqref{eq:parametrizationBar} and~\eqref{eq:solutionBar},
  \begin{eqnarray}
    \begin{bmatrix}
      \mathcal{C}_{x,y}^0(m,p)\\
      \mathcal{C}_{x,y}^1(m,p)\\
      \mathcal{C}_{x,y}^2(m,p)
    \end{bmatrix} = 4^x
    \begin{bmatrix}
      \bar{f}_{y-1}^0(m,p)& \bar{f}_{y-1}^1(m,p)& \bar{f}_{y-1}^2(m,p)\\
      \bar{g}_{y-1}^0(m,p)& \bar{g}_{y-1}^1(m,p)& \bar{g}_{y-1}^2(m,p)\\
      \bar{h}_{y-1}^0(m,p)& \bar{h}_{y-1}^1(m,p)& \bar{h}_{y-1}^2(m,p)
    \end{bmatrix}
    \begin{bmatrix}
      1 \\ 2 \\ 1
    \end{bmatrix}.
  \end{eqnarray}
  Similarly, the second part is
  \begin{eqnarray}
    \eqalign{
      \bra{c(x,t)}\overline{T}^{y}(1-Q)&=
      \sum_{m,p}\mathcal{D}_{x,y}^0(m,p)\bra{x-m,t-(x+y+1)-p;1;0}\\
      &+\sum_{m,p}\mathcal{D}_{x,y}^1(m,p)\bra{x-m,t-(x+y+1)-p;1;1}\\
      &+\sum_{m,p}\mathcal{D}_{x,y}^2(m,p)\bra{x-m,t-(x+y+1)-p;1;2},
    }
  \end{eqnarray}
  with
  \begin{eqnarray}\label{eq:coefficientsD}\fl
    \begin{bmatrix}
      \mathcal{D}_{x,y}^0(m,p)\\
      \mathcal{D}_{x,y}^1(m,p)\\
      \mathcal{D}_{x,y}^2(m,p)
    \end{bmatrix} =
    \sum_{c,w}
    \begin{bmatrix}
      \bar{f}_{y}^0(c,w)& \bar{f}_{y}^1(c,w)& \bar{f}_{y}^2(c,w)\\
      \bar{g}_{y}^0(c,w)& \bar{g}_{y}^1(c,w)& \bar{g}_{y}^2(c,w)\\
      \bar{h}_{y}^0(c,w)& \bar{h}_{y}^1(c,w)& \bar{h}_{y}^2(c,w)
    \end{bmatrix}
    \begin{bmatrix}
      -\mathcal{B}_x^0(m-c,p-w)\\
      \mathcal{B}_x^1(m-c,p-w)\\
      \mathcal{B}_x^2(m-c,p-w)
    \end{bmatrix}.
  \end{eqnarray}

  \subsection{The explicit overlaps $\Delta C_{l,r}(x,t)$}\label{subs:differentContributions}
  To express the overlaps~\eqref{eq:dClFormalSum} and~\eqref{eq:dCrFormalSum},
  we group the contributions from the different coefficients into separate groups,
  \begin{eqnarray}
    \eqalign{
      \Delta C_l(x,t)\equiv \Delta a(x,t) +\Delta b(x,t) +\Delta c(x,t) +\Delta d(x,t),\\
      \Delta C_r(x,t)\equiv \Delta a^{\prime}(x,t) +\Delta b^{\prime}(x,t) +\Delta c^{\prime}(x,t) +\Delta d^{\prime}(x,t),
    }
  \end{eqnarray}
  where $\Delta a(x,t)$ and $\Delta a^{\prime}(x,t)$ 
  include all the contributions from $\mathcal{A}_x^n$ coefficients,
  \begin{eqnarray}\fl \label{eq:DA}
    \eqalign{
      \Delta a(x,t)&=4^{t-x-1}\bigg(-\mathcal{A}_{x}^1(x-2,t-x-3)+\mathcal{A}_x^1(x-1,t-x-2)
      +\\&+2\mathcal{A}_x^1(x-1,t-x-3)-2\mathcal{A}_x^1(x-2,t-x-2)\bigg),\\
      \Delta a^{\prime}(x,t) &=4^{t-x-1}\bigg(
        +\mathcal{A}_x^0(x,t-x-1)+\frac{1}{2}\mathcal{A}_x^0(x,t-x-2)
        -\mathcal{A}_x^0(x-1,t-x-2)
        -\\&-\mathcal{A}_x^1(x,t-x-1)+2\mathcal{A}_x^1(x-1,t-x-1)
        -\frac{3}{2}\mathcal{A}_x^1(x,t-x-2)
      \bigg),
    }
  \end{eqnarray}
  $\Delta b(x,t)$ and $\Delta b^{\prime}(x,t)$ include the contributions of $\mathcal{B}_{x}^n$,
  \begin{eqnarray}\fl \label{eq:DB}
    \eqalign{
      \Delta b(x,t)&=4^{t-x-2}\bigg(
        -4\mathcal{B}^0_x(x,t-x-2)-\mathcal{B}_x^0(x,t-x-3)-\mathcal{B}_x^0(x-1,t-x-3)\\
        &+4\mathcal{B}_x^1(x-1,t-x-2)+3\mathcal{B}^1_x(x-1,t-x-3)\\
        &+2\mathcal{B}_x^2(x,t-x-2)+2\mathcal{B}_x^2(x-1,t-x-2)+\mathcal{B}_x^2(x-1,t-x-3)
      \bigg),\\
      \Delta b^{\prime}(x,t)  &=4^{t-x-2}\bigg(
        -2\mathcal{B}_x^0(x,t-x-2)-\frac{1}{2}\mathcal{B}_x^0(x,t-x-3)+\frac{3}{2}\mathcal{B}_x^1(x-1,t-x-3)
        +\\&+\mathcal{B}_x^2(x,t-x-2)+\frac{1}{2}\mathcal{B}_x^2(x-1,t-x-3)
      \bigg),
    }
  \end{eqnarray}
  $\Delta c(x,t)$ and $\Delta c^{\prime}(x,t)$ contain the contributions
  from $\mathcal{C}_{x,y}^n$,
  \begin{eqnarray} \label{eq:DC}
    \fl\eqalign{
      \Delta c(x,t) &= \sum_{y=1}^{t-x-2}
      4^{t-x-y-2}\bigg(
        4\mathcal{C}_{x,y}^0(x,t-x-y-2)+\mathcal{C}_{x,y}^0(x,t-x-y-3)
        +\\&+\mathcal{C}_{x,y}^0(x-1,t-x-y-3)
        +4\mathcal{C}_{x,y}^1(x-1,t-x-y-2)
        +\\&+3\mathcal{C}_{x,y}^1(x-1,t-x-y-3)+
        2\mathcal{C}_{x,y}^1(x,t-x-y-2)
        +\\&+2\mathcal{C}_{x,y}^2(x-1,t-x-y-2)+\mathcal{C}_{x,y}^2(x-1,t-x-y-3)
      \bigg),\\
      \Delta c^{\prime}(x,t)&=
      \sum_{y=1}^{t-x-2}
      4^{t - x - y - 2} \bigg(
        2 \mathcal{C}_{x,y}^0\left(x, t - x - y - 2\right) + 
        \frac{1}{2} \mathcal{C}_{x,y}^0\left(x, t - x - y - 3\right)  + \\ &+
        \frac{3}{2} \mathcal{C}_{x,y}^1\left(x - 1, t - x - y - 3\right) +
        \mathcal{C}_{x,y}^2\left(x, t - x - y - 2\right) + \\&+
        \frac{1}{2} \mathcal{C}_{x,y}^2\left(x - 1, t - x - y - 3\right)
      \bigg),
    }
  \end{eqnarray}
  and $\Delta d(x,t)$, $\Delta d^{\prime}(x,t)$ contain the contributions from $\mathcal{D}_{x,y}^n$,
  \begin{eqnarray}\fl \label{eq:DD}
    \eqalign{
      \Delta d(x,t) &= \sum_{y=1}^{t-x-2}
      4^{t-x-y-2}\bigg(
        4\mathcal{D}_{x,y}^0(x,t-x-y-2)+\mathcal{D}_{x,y}^0(x,t-x-y-3)
        +\\&+\mathcal{D}_{x,y}^0(x-1,t-x-y-3)
        +4\mathcal{D}_{x,y}^1(x-1,t-x-y-2)
        +\\&+3\mathcal{D}_{x,y}^1(x-1,t-x-y-3)
        + 2\mathcal{D}_{x,y}^2(x,t-x-y-2)
        +\\&+2\mathcal{D}_{x,y}^2(x-1,t-x-y-2)+\mathcal{D}_{x,y}^2(x-1,t-x-y-3)
      \bigg),\\
      \Delta d^{\prime}(x,t)&=
      \sum_{y=1}^{t-x-2}
      4^{t - x - y - 2} \bigg(
        2 \mathcal{D}_{x,y}^0\left(x, t - x - y - 2\right) + 
        \frac{1}{2} \mathcal{D}_{x,y}^0\left(x, t - x - y - 3\right)  + \\&+
        \frac{3}{2} \mathcal{D}_{x,y}^1\left(x - 1, t - x - y - 3\right) +
        \mathcal{D}_{x,y}^2\left(x, t - x - y - 2\right) + \\ &+
        \frac{1}{2}\mathcal{D}_{x,y}^2\left(x - 1, t - x - y - 3\right)
      \bigg).
    }
  \end{eqnarray}

  The contributions $\Delta a(x,t)$, $\Delta b(x,t)$, $\Delta a^{\prime}(x,t)$ and $\Delta b^{\prime}(x,t)$
  can be expressed in terms of simple binomial coefficients by plugging the coefficients
  $\mathcal{A}_x^n$, $\mathcal{B}_x^n$ into equations~\eqref{eq:DA} and~\eqref{eq:DB},
  \begin{eqnarray}\fl \label{eq:DAFinal}
    \eqalign{
      \Delta a(x,t)+\Delta b(x,t) &= \theta_{2x-t-2} 2^{3t-2x-3}
      \left(\binom{2x-t-2}{t-x-3}-\binom{2x-t-2}{t-x-2}\right)-\\
      &-\theta_{2x-t+1} 2^{3t-2x-7}
      \left(\binom{2x-t+1}{t-x-4}-\binom{2x-t+1}{t-x-3}\right),\\
      \Delta a^{\prime}(x,t)+\Delta b^{\prime}(x,t)  &=
      \theta_{2x-t-2} \left(2\binom{2x-t-2}{t-x-1}-\binom{2x-t-1}{t-x-3}\right)-\\
      &-\theta_{2x-t+1} 2^{3t-2x-8}\left(\binom{2x-t+1}{t-x-3}-\binom{2x-t+1}{t-x-4}\right).
    }
  \end{eqnarray}
  Similarly, the sums~\eqref{eq:DC} can be simplified into the following form,
  \begin{eqnarray} \fl \label{eq:DCFinal}
    \eqalign{
      \Delta c(x,t)=\theta_{t-2x-4} 2^{t+2x-1}\left(
        6\binom{t-2x-4}{x}+5\binom{t-2x-4}{x-1}+\binom{t-2x-4}{x-2}
      \right),\\
      \Delta c(x,t)^{\prime}=\theta_{t-2x-4} 2^{t+2x-1}\left(
        3\binom{t-2x-4}{x}+\binom{t-2x-4}{x-1}
      \right),
    }
  \end{eqnarray}
  by observing that for any $u\ge 0$ the following holds,
  \begin{eqnarray}\fl \label{eq:sumSDefinition}
    \sum_{m=0}^{\lfloor\frac{u}{2}\rfloor} 4^m\binom{u-2m}{m}=2^{u-1}
    +\frac{1-\frac{\ii}{\sqrt{7}}}{4}\left(\frac{-1+\ii\sqrt{7}}{2}\right)^{\!u}\!\!
    +\frac{1+\frac{\ii}{\sqrt{7}}}{4}\left(\frac{-1-\ii\sqrt{7}}{2}\right)^{\!u}\!\!\equiv a_u.
  \end{eqnarray}
  However, simplifying the contributions $\Delta d(x,t)$ and $\Delta d^{\prime}(x,t)$
  requires a bit more work.

  \subsection{Contributions \texorpdfstring{$\Delta d(x,t)$}{Δd(x,t)} and \texorpdfstring{$\Delta d^{\prime}(x,t)$}{Δ'd(x,t)}}
  \label{subs:deltaDdeltaDprime}
  We start by noting that both the remaining contributions
  can be expressed in terms of the following triple sum,
  \begin{eqnarray}\fl\label{eq:DefSxt}
    \eqalign{
      s_{x,t}(\alpha,\beta,\gamma)=
      \smashoperator[l]{\sum_{z=0}^{\min\{x+\alpha,t-x+\beta\}}}
      \smashoperator[r]{\sum_{y=0}^{\min\{x+\alpha,t-x+\beta\}-z}}
      2^{-(x+\alpha-y-z)} \binom{x+\alpha-y-z}{y}\times\\ \times
      2^{-(t-x+\beta-y-z)} \binom{t-x+\beta-y-z}{z}
      \quad\ \smashoperator{\sum_{w=0}^{\frac{1}{2}(t-x+\gamma-y+z)}}
      2^{-(t-x+\gamma-y+z-2w)}\binom{t-x+\gamma-y+x-2w}{w},
    }
  \end{eqnarray}
  as
  \begin{eqnarray}\fl\label{eq:DDSumFull}
    \eqalign{
      \Delta d(x,t) &- 2\Delta d^{\prime}(x,t)=
      2^{3t-2x-6}\binom{2x-t+2}{t-x-3}
      +\\&+2^{2t-5} \bigg(
        s_{x,t}( -3, -4, -3)
        + s_{x,t}( -3, -4, -2)
        + 2 s_{x,t}( -3, -4, -1)
        +\\&+ s_{x,t}( -2, -3, -6)
        + s_{x,t}( -2, -3, -5) 
        + 2 s_{x,t}( -2, -3, -4)
        -\\&- 2 s_{x,t}( -1, -3, -4) 
        - 2 s_{x,t}( -1, -3, -3) 
        - 4 s_{x,t}( -1, -3, -2)
      \bigg),
    }
  \end{eqnarray}
  and
  \begin{eqnarray}\fl\label{eq:DDpFull}
    \eqalign{
      \Delta d^{\prime}(x,t) &=
      2^{3t-2x-8} \left(8 \binom{2 x - t + 2}{t - x - 2} + 
      \binom{2 x - t + 3}{t - x - 3}\right)+\\
      &+2^{2 t-7} \bigg(
        s_{x,t}(-3,-5,-4) 
        + 3 s_{x,t}(-3,-5,-2) 
        + s_{x,t}(-2,-4,-7)
        +\\&+ 3 s_{x,t}(-2,-4,-5) 
        + 2 s_{x,t}(-2,-4,-3) 
        + 2 s_{x,t}(-2,-4,-2)
        +\\&+ 8 s_{x,t}(-2,-3,-1) 
        - 2 s_{x,t}(-1,-4,-5) 
        - 6 s_{x,t}(-1,-4,-3)
        +\\&+ 2 s_{x,t}(-1,-3,-6) 
        + 2 s_{x,t}(-1,-3,-5) 
        + 8 s_{x,t}(-1,-2,-4)
        -\\&- 4 s_{x,t}(0,-3,-4) 
        - 4 s_{x,t}(0,-3,-3) 
        -16 s_{x,t}(0,-2,-2)
      \bigg).
    }
  \end{eqnarray}
  Note that instead of explicitly expressing $\Delta d(x,t)$ we simplified the expressions a bit
  by introducing $\Delta d(x,t)-2\Delta d^{\prime}(x,t)$.

  We first observe that the inner-most sum 
  in~\eqref{eq:DefSxt} can be evaluated,
  \begin{eqnarray}
    \sum_{w=0}^{\frac{u}{2}}2^{-u+2w}\binom{u-2w}{w}= 2^{-u}a_{u},
  \end{eqnarray}
  with $a_u$ defined in~\eqref{eq:sumSDefinition}. If $u\ge0$, the coefficients $a_u$ satisfy the following
  recurrence relation,
  \begin{eqnarray}
    a_{u+1}=a_u+4 a_{u-2}. 
  \end{eqnarray}
  This enables us to rewrite the expression~\eqref{eq:DDSumFull} in terms of simpler double sums,
  \begin{eqnarray}
    \bar{s}(m,n)=\smashoperator[l]{\sum_{z=0}^{\min\{m,n\}}}\smashoperator[r]{\sum_{y=0}^{\min\{m,n\}-z}}
    4^{y+z}\binom{m-y-z}{y}\binom{n-y-z}{z},
  \end{eqnarray}
  by grouping together the terms $s_{x,t}(\alpha,\beta,\gamma)$ with the same $\alpha$, $\beta$.
  Explicitly,
  \begin{eqnarray}\fl
    s_{x,t}(\alpha,\beta,\gamma)+s_{x,t}(\alpha,\beta,\gamma+1)+2 s_{x,t}(\alpha,\beta,\gamma+2)
    \mapsto
    2^{-(t+\alpha+\beta)}\bar{s}(x+\alpha,t-x+\beta).
  \end{eqnarray}
  Furthermore, we have to subtract the terms that contain $a_{n}$ with $n<0$, since the relation does not
  hold for them. Taking care of these corner cases,
  the contribution $\Delta d(x,t)-2\Delta d^{\prime}(x,t)$ can be rewritten as
  \begin{eqnarray}\fl
    \eqalign{
      &\Delta d(x,t)-2\Delta d^{\prime}(x,t)=2^{3t-2x-6}\binom{2x-t+1}{t-x-3}-2^{3t-2x-5}\binom{2x-t+1}{t-x-3}
      +\\ &+ 2^{t+3} \bar{s}(x-3,t-x-4)+2^{t+1}\bar{s}(x-2,t-x-3)-2^{t+1}\bar{s}(x-1,t-x-3).
    }
  \end{eqnarray}
  It is possible to further simplify this result by noting that
  the functions $\bar{s}(m,n)$ satisfy the following two relations,
  \begin{eqnarray}\fl
    \eqalign{
      \bar{s}(m+1,n)=\bar{s}(m,n)+4\bar{s}(m-1,n-1)+\varepsilon(m,n),\\
      \bar{s}(m,n+1)=\bar{s}(m,n)+4\bar{s}(m-1,n-1)+\eta(m,n),\\
      \varepsilon(m,n)=\theta_{n-m-1} 4^{m+1}\binom{n-m-1}{m+1},\quad
      \eta(m,n)=\theta_{m-n-1} 4^{n+1}\binom{m-n-1}{n+1},
    }
  \end{eqnarray}
  which enable us to obtain
  \begin{eqnarray}\label{eq:DDSumFinal}\fl
    \Delta d(x,t)-2\Delta d^{\prime}(x,t)=
    \begin{cases}
      -2^{2x+t-1}\binom{t-2x-2}{x-1}; & x\le \frac{t-2}{2},\\
      -2^{3t-2x-6}\left(
        \binom{2x-t+1}{t-x-3}-\binom{2x-t+1}{t-x-4}
      \right); & x \ge \frac{t-1}{2}.
    \end{cases}
  \end{eqnarray}

  The contribution~\eqref{eq:DDpFull} is a bit more complicated, since the sums $s_{x,t}(\alpha,\beta,\gamma)$
  with the same $\alpha$, $\beta$ do not simplify as before. Therefore we split
  the coefficients $a_n$ into two parts,
  \begin{eqnarray}
    2^{-n}a_n=\frac{1}{2}+b_n,
  \end{eqnarray}
  and we treat the different contributions to $\Delta d^{\prime}(x,t)$ separately,
  \begin{eqnarray}
    \Delta d^{\prime}(x,t) = \Delta d_c^{\prime}(x,t) + \Delta d_r^{\prime}(x,t) + \Delta d_i^{\prime}(x,t),
  \end{eqnarray}
  where $\Delta d_c^{\prime}(x,t)$ includes all the constant terms,
  \begin{eqnarray}
    \Delta d_c^{\prime}(x,t) = 2^{3t-2x-8}\left(
      8\binom{2x-t+2}{t-x-2}
      -\binom{2x-t+2}{t-x-3}
    \right),
  \end{eqnarray}
  and $\Delta d_{i,r}^{\prime}(x,t)$ correspond to different parts of the coefficients $a_n$. Explicitly,
  \begin{eqnarray}\label{eq:defDDpr}
    \fl\eqalign{
      \Delta d^{\prime}_r(x,t)&=2^{t-2}\bigg(
        16\bar{s}(x-3,t-x-5)+8\bar{s}(x-2,t-x-4)+4\bar{s}(x-2,t-x-3)
        -\\&-4\bar{s}(x-1,t-x-4)
        +\bar{s}(x-1,t-x-3)+\bar{s}(x-1,t-x-2)
        -\\&-\bar{s}(x,t-x-3)-\bar{s}(x,t-x-2)
      \bigg),
    }
  \end{eqnarray}
  and
  \begin{eqnarray}\label{eq:defDDpi}
    \fl \eqalign{
      \Delta d^{\prime}_i(x,t) &=-2^{t-1} \bigg(
        4\bar{s}_2(x-3,t-x-5)
        - 4\bar{s}_3(x-3,t-x-5)
        +\\&+ 2\bar{s}_1(x-2,t-x-4)
        - 2\bar{s}_2(x-2,t-x-4)
        - 4\bar{s}_2(x-2,t-x-3)
        -\\&-  \bar{s}_0(x-1,t-x-4)
        +  \bar{s}_1(x-1,t-x-4)
        -  \bar{s}_0(x-1,t-x-3)
        -\\&- 2\bar{s}_1(x-1,t-x-3)
        +  \bar{s}_0(x-1,t-x-2)
        - 2\bar{s}_1(x-1,t-x-2)
        -\\&-  \bar{s}_1(x,t-x-3)
        +  \bar{s}_0(x,t-x-2)
      \bigg),
    }
  \end{eqnarray}
  with the generalized sums $\bar{s}_\gamma(m,n)$ defined as
  \begin{eqnarray}\label{eq:barsgammadef}
    \bar{s}_{\gamma}(m,n) =
    \smashoperator[l]{\sum_{z=0}^{\min\{m,n\}}}
    \smashoperator[r]{\sum_{y=0}^{\min\{m,n\}-z}}
    4^{y+z}\binom{m-y-z}{y}\binom{n-y-z}{z}b_{n+1+\gamma-y-z}.
  \end{eqnarray}
  Similarly as before, the contribution~\eqref{eq:defDDpr} reduces into
  \begin{eqnarray}\fl
    \eqalign{
      \Delta d^{\prime}_r(x,t)&=2^{t-2}\bigg(
        -4\eta(x-2,t-x-4)-\eta(x-1,t-x-3)+\eta(x,t-x-3)
        -\\&-2\varepsilon(x-1,t-x-2)
      \bigg)
      =\\&=-\theta_{t-2x-2}2^{2x+t-1}\binom{t-2x-2}{x}.
    }
  \end{eqnarray}
  In order to simplify the last part~\eqref{eq:defDDpi}, we first observe that the following
  relations hold,
  \begin{eqnarray}\label{eq:sbargammaRec}
    \eqalign{
      \bar{s}_\gamma(m+1,n)
      = \bar{s}_{\gamma}(m,n) + 4\bar{s}_{\gamma}(m-1,n-1)
      + \varepsilon_{{\gamma}}(m,n),\\
      \bar{s}_{\gamma}(m,n+1) = \bar{s}_{\gamma+1}(m,n)
      + 4 \bar{s}_{\gamma+3}(m-1,n-1) + \eta_{{\gamma}}(m,n),\\
      \varepsilon_{{\gamma}}(m,n)= \theta_{n-m-1}
      4^{m+1}\binom{n-(m+1)}{m+1} b_{n+m+1+\gamma},\\
      \eta_{{\gamma}}(m,n)=\theta_{m-n-1}
      4^{n+1}\binom{m-(n+1)}{n+1}b_{\gamma}.
    }
  \end{eqnarray}
  Using them, we obtain,
  \begin{eqnarray}\fl
    \eqalign{
      \Delta d^{\prime}_{i}(x,t)&=-2^{t-1}\bigg(
        \eta_{0}(x,t-x-3)-\eta_1(x-1,t-x-3)-4\eta_2(x-2,t-x-4)
      +\\&+4\eta_3(x-1,t-x-3)\bigg)
      =\\&=-\theta_{2x-t+1} 2^{3t-2x-7}\left(
        3\binom{2x-t+1}{t-x-3}+4\binom{2x-t+1}{t-x-2}
      \right),
    }
  \end{eqnarray}
  which yields
  \begin{eqnarray}\label{eq:DDpFinal}
    \Delta d^{\prime}(x,t)=\begin{cases}
      -2^{2x+t-1}\binom{t-2x-2}{x}; & x\le \frac{t-2}{2},\\
      2^{3t-2x-9}\left(\binom{2x-t+1}{t-x-3}-\binom{2x-t+1}{t-x-4}\right); & x\ge\frac{t-1}{2}.
    \end{cases}
  \end{eqnarray}

  By combining the equations~\eqref{eq:DAFinal}, \eqref{eq:DCFinal}, \eqref{eq:DDSumFinal} and~\eqref{eq:DDpFinal},
  we can finally express the left and right overlap as
  \begin{eqnarray} \fl 
    \eqalign{
      \Delta C_l(x,t)=\begin{cases}
        2^{2x+t-1}\left(
          4\binom{t-2x-4}{x}
          -3\binom{t-2x-4}{x-2}
          -\binom{t-2x-4}{x-3}
        \right),& x\le \frac{t-4}{2},\\
        0,& \frac{t-3}{2}\le x \le \frac{t+1}{2},\\
        2^{3t-2x-3}\left(
          \binom{2x-t-2}{t-x-3}-\binom{2x-t-2}{t-x-2}
        \right), & \frac{t+2}{2}\le x,
      \end{cases}\\
      \Delta C_r(x,t)=\begin{cases}
        2^{2x+t-1}\left(
          2\binom{t-2x-4}{x}
          -\binom{t-2x-4}{x-1}
          -\binom{t-2x-4}{x-2}
        \right),& x\le\frac{t-4}{2},\\
        0,& \frac{t-3}{2}\le x \le \frac{t+1}{2},\\
        2^{3t-2x-3}\left(
          2\binom{2x-t-2}{t-x-1}
          -\binom{2x-t-2}{t-x-3}
          -\binom{2x-t-2}{t-x-4}
        \right),& \frac{t+2}{2}\le x.
      \end{cases}
    }
  \end{eqnarray}
  \subsection{The equation~\texorpdfstring{\eqref{eq:STcorrsPoly}}{placeholder}}\label{sec:STcorrsPoly}
  To show that~\eqref{eq:STcorrsPoly} is equivalent to~\eqref{eq:formalSumSpatioTemporal}, it suffices
  to prove that the polynomial $\tilde{p}(u,x)$, defined as
  \begin{eqnarray}
    \tilde{p}(u,x) = \sum_{n=2x}^{3x}\left(\prod_{\substack{j=2x\\j\neq n}}^{3x}\frac{u-j}{n-j}\right)\left(2 s(n)-s(n+1)\right),
  \end{eqnarray}
  is equivalent to the sum $p(u,x)$,
  \begin{eqnarray}
    p(u,x) = \sum_{m=0}^{x}4^m\left(
      2\binom{u-2m}{m}
      -\binom{u+1-2m}{m}
    \right),
  \end{eqnarray}
  where $s(u)=p(u,\lfloor\frac{u}{2}\rfloor)$ was defined in the main text~\footnote{This is also
  the same as $a_u$, defined in~\eqref{eq:sumSDefinition}.} and $u\ge0$, $2x \le u$. Clearly,
  if $\frac{u}{2}\ge x \ge \frac{u+1}{3}$, both expressions coincide, therefore it is sufficient to
  show that $\tilde{p}(u,x)$ satisfies the same relation as $p(u,x)$,
  \begin{eqnarray}\fl
    p(u,x)=p(u-1,x)+4p(u-2,x)-4^{x+1}\left(
      2\binom{u-2x-3}{x}
      -\binom{u-2x-2}{x}
    \right).
  \end{eqnarray}
  After some straightforward manipulation of the sums, we obtain the following
  \begin{eqnarray}
    \fl \eqalign{
      \tilde{p}(u,x)-\tilde{p}(u-1,x)-4\tilde{p}(u-3,x)=\\
      =\smashoperator{\sum_{n=2x}^{3x}}\left(\prod_{\substack{j=2x\\j\neq n}}^{3x}\frac{u-j}{n-j}\right)
      \overbrace{\left(
          -s(n+1)+3 s(n)-2 s(n-1)+4 s(n-2)-8 s(n-3)
      \right)}^{=0}
      +\\+\smashoperator{\sum_{n=2x+2}^{3x+3}} (-1)^{n+3x}\binom{x+1}{n-2x-2}\binom{u-2x-3}{x}\left(8 s(n-3)-4 s(n-2)\right)
      +\\+\smashoperator{\sum_{n=2x+1}^{3x+2}}(-1)^{n+3x}\binom{x+1}{n-2x-1}\binom{u-2x-2}{x}\left(8 s (n-3)-4 s(n-2)\right)
      +\\+\smashoperator{\sum_{n=2x}^{3x+1}} (-1)^{n+3x}\binom{x+1}{n-2x}\binom{u-2x-1}{x}
      \overbrace{\left(-s(n)+2 s(n-1)-4 s(n-2)+8 s(n-3)\right)}^{=2 s(n)-s(n+1)}.
    }
  \end{eqnarray}
  Expressing it in terms of
  the sums $r(x,\alpha)=\sum_{n=0}^{x+1}(-1)^n\binom{x+1}{n}s(n+\alpha)$ and
  taking into account the following properties,
  \begin{eqnarray}
    \left.
      \eqalign{
        r(x+1,\alpha)=-4 r(x,\alpha-2)\\
        r(0,\alpha)=-4 s(\alpha-2)
    }\right\}\,
    r(x,\alpha)=(-4)^{x+1} s(\alpha-2x-2),
  \end{eqnarray}
  yields
  \begin{eqnarray}\fl
    \tilde{p}(u,x)\!-\!\tilde{p}(u-1,x)\!-\!4\tilde{p}(u-2,x)=
    \!-4^{x+1}\!\left(
      \!
      2\binom{u-2x-3}{x}
      -\binom{u-2x-2}{x}
    \!\right).
  \end{eqnarray}
  \end{document}